\documentclass[floatfix,amsmath,amssymb,aps,
prb,superscriptaddress,natbib,twocolumn]{revtex4-2}
\usepackage{graphicx}
\usepackage{dcolumn}
\usepackage{bm}
\usepackage[utf8]{inputenc}
\usepackage{chemformula}
\usepackage[caption=false]{subfig}

\DeclareUnicodeCharacter{2212}{-} 

\def\YBCO{YBa$_2$Cu$_3$O$_{7-\delta}$}

\def\C60{A$_x$C$_{60}$}

\def\SROtwo{ Sr$_{3}$Ru$_{2}$O$_{7}$}

\def\HgCu3{HgCa$_2$Cu$_3$O$_{8+y}$}
\def\HgCu4{HgBa$_2$Ca$_3$Cu$_4$O$_{10+y}$}
\def\TlCu{Tl$_2$Ba$_2$CuO$_{6+\delta}$}
\def\TlCu3{Tl$_2$Ba$_2$Ca$_2$Cu$_3$O$_{10+y}$}
\def\TlCu4{Tl$_2$Ba$_2$Ca$_3$Cu$_4$O$_{12+y}$}

\def\BiCu3{Bi$_2$Sr$_2$Ca$_{2}$Cu$_3$O$_y$}

\def\8LSCO{La$_{1.88}$Sr$_{.12}$CuO$_4$}
\def\110LNSCO{La$_{1.5}$Nd$_{0.4}$Sr$_{0.1}$CuO$_{4}$}
\def\stage4LCO{La$_{2}$CuO$_{4+\delta}$}
\def\Y248{YBa$_2$Cu$_4$O$_8$}

\def\NbSe2{NbSe$_2$}
\def\TaSe2{TaSe$_2$}
\def\TiSe2{TiSe$_2$}
\def\NaCoOH2O{Na$_{0.3}$CoO$_{2y}$H$_2$O}
\def\MgB2{MgB${}_2$}

\def\BaFeCoAs{Ba(Fe$_{1-x}$Co$_x$)$_2$As$_2$}
\def\BaSrNiAs{Ba$_{1-x}$Sr$_x$Ni$_2$As$_2$
}

\begin{document}

\title{A Ginzburg-Landau approach to the vortex-domain wall interaction in superconductors with nematic order }
\author{R. S. Severino}
\affiliation{Universidad de Buenos Aires, Facultad de Ciencias Exactas y Naturales, Departamento de Física. Buenos Aires, Argentina.}
 \affiliation{CONICET - Universidad de Buenos Aires, Instituto de Física de Buenos Aires (IFIBA). Buenos Aires, Argentina}
\author{P. D. Mininni}
\affiliation{Universidad de Buenos Aires, Facultad de Ciencias Exactas y Naturales, Departamento de Física. Buenos Aires, Argentina.}
\affiliation{CONICET - Universidad de Buenos Aires, Instituto de Física Interdisciplinaria y Aplicada (INFINA). Buenos Aires, Argentina}
\author{E. Fradkin}
 \affiliation{Department of Physics and Institute for Condensed Matter Theory, University of Illinois  at Urbana-Champaign,
 1110 West Green Street, Urbana, Illinois 61801-3080, USA}
 \author{V. Bekeris}
\affiliation{Universidad de Buenos Aires, Facultad de Ciencias Exactas y Naturales, Departamento de Física. Buenos Aires, Argentina.}
 \affiliation{CONICET - Universidad de Buenos Aires, Instituto de Física de Buenos Aires (IFIBA). Buenos Aires, Argentina}
 \author{G. Pasquini}
\affiliation{Universidad de Buenos Aires, Facultad de Ciencias Exactas y Naturales, Departamento de Física. Buenos Aires, Argentina.}
\affiliation{CONICET - Universidad de Buenos Aires, Instituto de Física de Buenos Aires (IFIBA). Buenos Aires, Argentina}
\author{G. S. Lozano}
\affiliation{Universidad de Buenos Aires, Facultad de Ciencias Exactas y Naturales, Departamento de Física. Buenos Aires, Argentina.}
 \affiliation{CONICET - Universidad de Buenos Aires, Instituto de Física de Buenos Aires (IFIBA). Buenos Aires, Argentina}
\begin{abstract}
In this work we study the interaction between vortices and nematic domain walls within the framework of a Ginzburg Landau approach. The free energy of the system is written in terms of a complex order parameter characteristic of $s$-wave superconductivity and a real  (Ising type) order  parameter associated to nematicity. The interaction between both order parameters is described by a biquadratic and a trilinear derivative term. To study the effects of these interactions we solve the time-dependent dissipative Ginzburg Landau equations using a highly performant pseudospectral method by which we calculate  the trajectories of a vortex that, for different coupling parameters, is either attracted or repelled by a wall, as well as of the wall dynamics. We show that despite its simplicity, this theory displays many phenomena observed experimentally in Fe-based superconductors. In particular we find that the sign of the biquadratic term determines 
the attractive (pining) or repulsive (antipining) character of the interaction, as observed in FeSe and BaFeCoAs compounds respectively. The trilinear term is responsible for the elliptical shape of vortex cores as well as for the orientation of the axes of the ellipses and vortex trajectories with respect to the axes of the structural lattice. For the case of pining, we show that the vortex core is well described by a heart-shaped structure in agreement with STM experiments performed in FeSe.
\end{abstract}
\maketitle

\section{Introduction}

The existence of the theoretically proposed electronic nematic phase in strongly correlated systems \cite{kivelson1, Fradkin_2010} has been confirmed through an amount of experimental work conducted in several systems \cite{Fradkin_2010}. The best documented examples of electronic nematicity are two-dimensional electron fluids in large magnetic fields at Landau levels $N \geq 1$ \cite{lilly-1999,kivelson2} and in the bi-layer ruthenate {\SROtwo} (in a range of magnetic fields) \cite{borzi-2007}. Nematicity  is also seen in diverse systems and it is believed to play a central role in  unconventional superconductors \cite{Fradkin_2015, Fernandes_2022}, in both cuprate \cite{Ando_2002, Hinkov_2008, Comin_2015} and iron-based \cite{Chuang_2010, Prozorov_2009, Kuo_2016, Chu_2010, Kuo_2012, Gallais_2013, Tanatar_2016, Kretzschmar_2016} family compounds. A key fact supporting the interplay between nematic and superconducting orders is the fact that their phase boundaries, as a function of doping, intersect near the composition that maximises  the superconducting critical temperature \cite{Prozorov_2009}, where the signature of a nematic quantum critical point has been reported \cite{Kuo_2016}. A similar behaviour has been reported recently in the family of nickel compounds {\BaSrNiAs} \cite{Eckberg-2020,Lee-2021}.

\textit{Electronic nematic order} \cite{Fradkin_2010} is a state of the electronic 
fluid that \textit{spontaneously} breaks the point group symmetry of the underlying lattice. In this paper we will consider the case of systems with translational symmetry along the $c$ axis and native $C_4$ point group symmetry of the $a-b$ plane spontaneously broken down to one of its $C_2$ subgroups, with symmetry $d_{x^2-y^2}$ or $d_{xy}$. Although this is the most common case, other point group symmetries, such as $C_6$, are relevant in several systems such as dichalcogenide \cite{Manzeli-2017} and kagome materials \cite{Wilson-2023}. Several microscopic mechanisms can give rise to an electronic nematic state including a Pomeranchuk instability \cite{oganesyan-2001,halboth-2000,metzner-2003}, orbital order \cite{kivelson-2004,Lv-2009}, or can arise as a vestigial order of a magnetically \cite{Xu-2008,Fang-2008} and/or charge ordered state \cite{Nie-2017,Fernandes-2019}. 
Electronic nematic states typically do not arise in weakly interacting systems and are the result of strong correlation effects.

In most experimental situations, the nematic phase manifests as the result of the breaking of a \textit{discrete} $C_4$ point-group symmetry of the underlying crystal structure of 
the material. It is generally expected that in a nematic state the crystal should distort since there is always a coupling between the electronic degrees of freedom and the lattice. How large this distortion is depends on the microscopic mechanism behind the nematic order. In some systems, such as in the underdoped {\YBCO} the orthorhombic crystal structure sets in at high temperatures and electron nematicity is observed as an ordering effect seen at lower temperatures \cite{Hinkov_2008}. On the other hand, in some pnictides, such as in {\BaFeCoAs}, nematic order (without long range magnetic order) arises on a relatively small fraction of the phase diagram while in FeSe this situation occurs in a broader range of temperatures and doping.

We should note that it is conceptually important to distinguish a tetragonal-to-orthorhombic structural transition from an electronic nematic transition even though both break the same point group symmetry. Most structural transitions are typically first order phase transitions and generally do not exhibit significant thermal ordering effects. In contrast, electronic nematic transitions can be continuous and have significant thermal fluctuations near the ordering transition. The nematic transition in the Fe superconductors appears to be either continuous or weakly first order, which strongly suggests that it is primarily driven by an electronic mechanism. The Curie Law divergence of the nematic susceptibility in the tetragonal phase is another evidence supporting its electronic origin \cite{Chu2012, Kuo_2016}. However, since the nematic state breaks the same symmetries as an orthorhombic crystal structure
it is hard to disentangle both effects. In particular, many experiments often 
show a concurrent breaking of the $C_4$ symmetry in the structural and transport properties consistent with a transition driven by electronic degrees of freedom \cite{Chu_2010, Kuo_2012, Gallais_2013, Tanatar_2016, Kretzschmar_2016}. On the other hand, despite the considerable amount of experimental evidence of the existence of electronic nematicity in the Fe superconductors, there is still no consensus on its microscopic origin.

One of the consequences of the appearance of a symmetry-breaking nematic phase is the formation of a 
dense array of structural and nematic domains (ND), that has been observed and characterised by several techniques \cite{Prozorov_2009, Kalisky_2010}. The role of these domains and the walls separating them, the so-called nematic domain walls (NDW),  in the properties of the normal metallic phase is still under intense study and debate \cite{sanches_nature, Bartlett_2021}. 

An interesting and less explored issue concerns the relevance of the nematic domains in the superconducting phase. For instance NDW will, in principle, interact with superconducting vortices. 
As nematicity is naturally coupled to anisotropic strain, the existence of nematic domains has been linked in some systems to the existence and location of \textit{structural twin boundaries} (TB), which have been identified in several works as sources of correlated vortex pinning (and in some cases of vortex channels) on their own \cite{Blatter1994, Kwok_1996, Crabtree_1996, Herbsommer_2000}. The influence of NDW/TB in the vortex dynamics in Fe based nematic superconductors has been reported in different magnetic and transport experiments \cite{Prozorov_2009, Marziali_2013, schmidt}, 
and the NDW/TB-vortex interaction has been directly observed in FeSe  \cite{song, songprl, Watashige_2015, zhang} and {\BaFeCoAs} \cite{yin, Kalisky_2011, Yagil_2016} compounds. While in the first case vortices are pinned by domain boundaries, in the second case vortices avoid them.

At a phenomenological level the interplay between nematicity and superconductivity can be in principle described via the Ginzburg-Landau (GL) formalism. In its simpler version, the GL free energy can be expressed as a functional of two order parameters, a complex field associated to superconductivity and a real (Ising type) field associated to nematicity. To lowest order, these fields are coupled via a trilinear term (involving derivatives) and a biquadratic term. The main effect of the trilinear term is to give rise to an anisotropy in the superfluid density of the superconducting state, and manifests in the elliptical shape of vortex cores. Although the coefficient of the biquadratic term has to satisfy certain bounds, its sign is not fixed a priori, representing order competition (cooperation) when it is positive (negative). Experimentally, the cooperative or competitive coupling manifests in several ways. For example, in-plane anisotropic superconducting properties are reflected in elongated superconducting vortices oriented along the crystallographic axes of the orthorhombic phase, as observed in early STM studies in FeSe superconductors \cite{song}. Also, flux flow resistivity anisotropy \cite{schmidt}, anisotropic critical current density \cite{Hecher_2018}, and anisotropic superconducting gap \cite{hashimoto} are additional evidences for in-plane broken symmetry in the superconducting properties.

Finally, it is worth mentioning that the coupling of electronic nematicity with strain might be more complex than generally believed \cite{ren}. However, even in a simpler scenario where NDW are univocally linked with structural TB they are an \textit{additional} source of pinning (or antipinning), in a way such that the final vortex-domain wall interaction will result from the combination of structural and nematic vortex interactions.

The purpose of this work is to analyse from a theoretical perspective some properties of the NDW-vortex interaction in the framework of a time-dependent GL theory. To this end, we will use the simplest GL theory of a nematic superconductor, with an $s$-wave superconductor order parameter and an Ising-like nematic order parameter. Nematic domain walls can be present for  two types of nematic orders with $d_{xy}$ and $d_{x^2-y^2}$ symmetry respectively, and the elliptically distorted vortices are oriented differently relative to the crystallographic axis. For definiteness, in the body of the paper we consider primarily the $d_{xy}$ nematic ordering. To further characterise the interaction we calculate trajectories, energies and forces as a function of time and distance. Despite the simplicity of the model, we find that many features are  qualitatively consistent with  experimental findings. The model can successfully capture the elongation of vortex cores in the presence of nematic ordering, and reproduces the pinning (antipinning) of vortices on domain walls as well as other features observed in experiments.  

This paper is organised as follows. In Section \ref{GL model} we describe the phenomenological GL model for the coupled superconducting and scalar nematic order parameters. The main calculations are presented in Section \ref{vortex} organised in different subsections. The numerical method is first described, followed by  the domain boundary effect in the superconducting order parameter and in the single vortex dynamics, where the  role of the biquadratic and trilinear coupling terms is considered. Finally, a summary and conclusions are presented in Section \ref{conclusions}. 

\section{Ginzburg-Landau Model}\label{GL model}

The  total Ginzburg-Landau (GL) free energy of interest is
\begin{equation}
    F=F_S+F_N+F_{SN} ,
    \label{F}
\end{equation}
\begin{figure}
    \subfloat[\label{fig:fig1a}]{%
      \includegraphics[width=0.4\columnwidth]{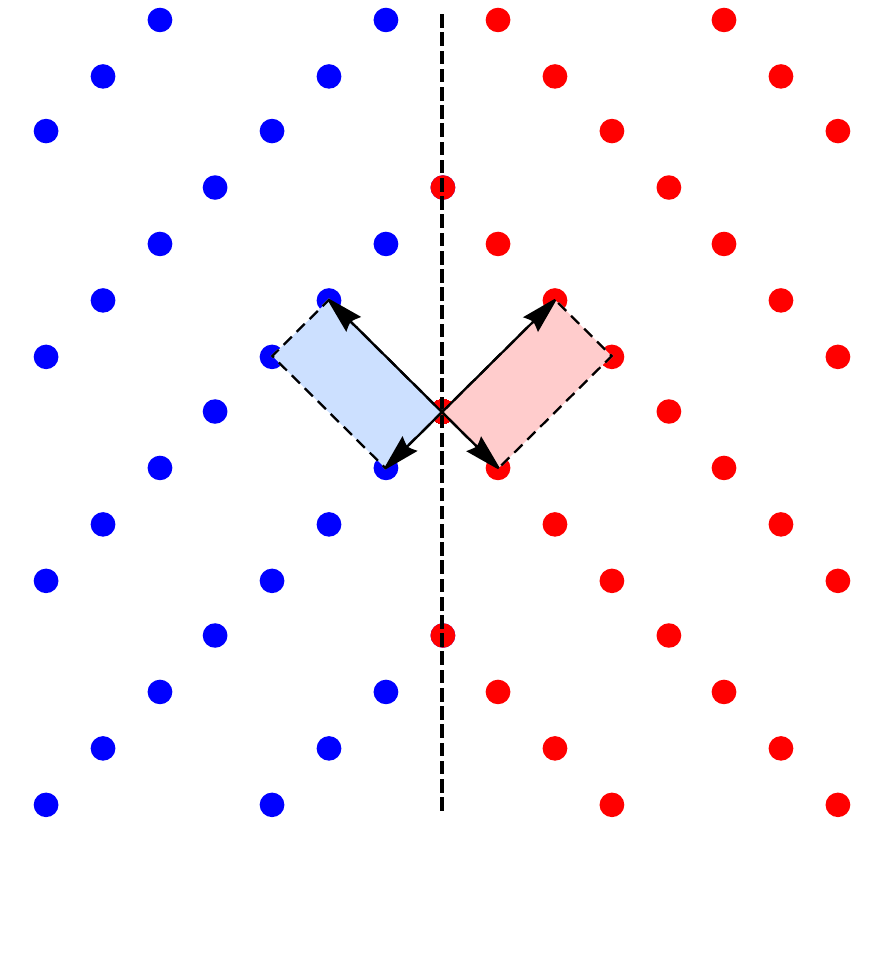}
    }
    \hfill
    \subfloat[\label{fig:fig1b}]{%
      \includegraphics[width=0.48\columnwidth]{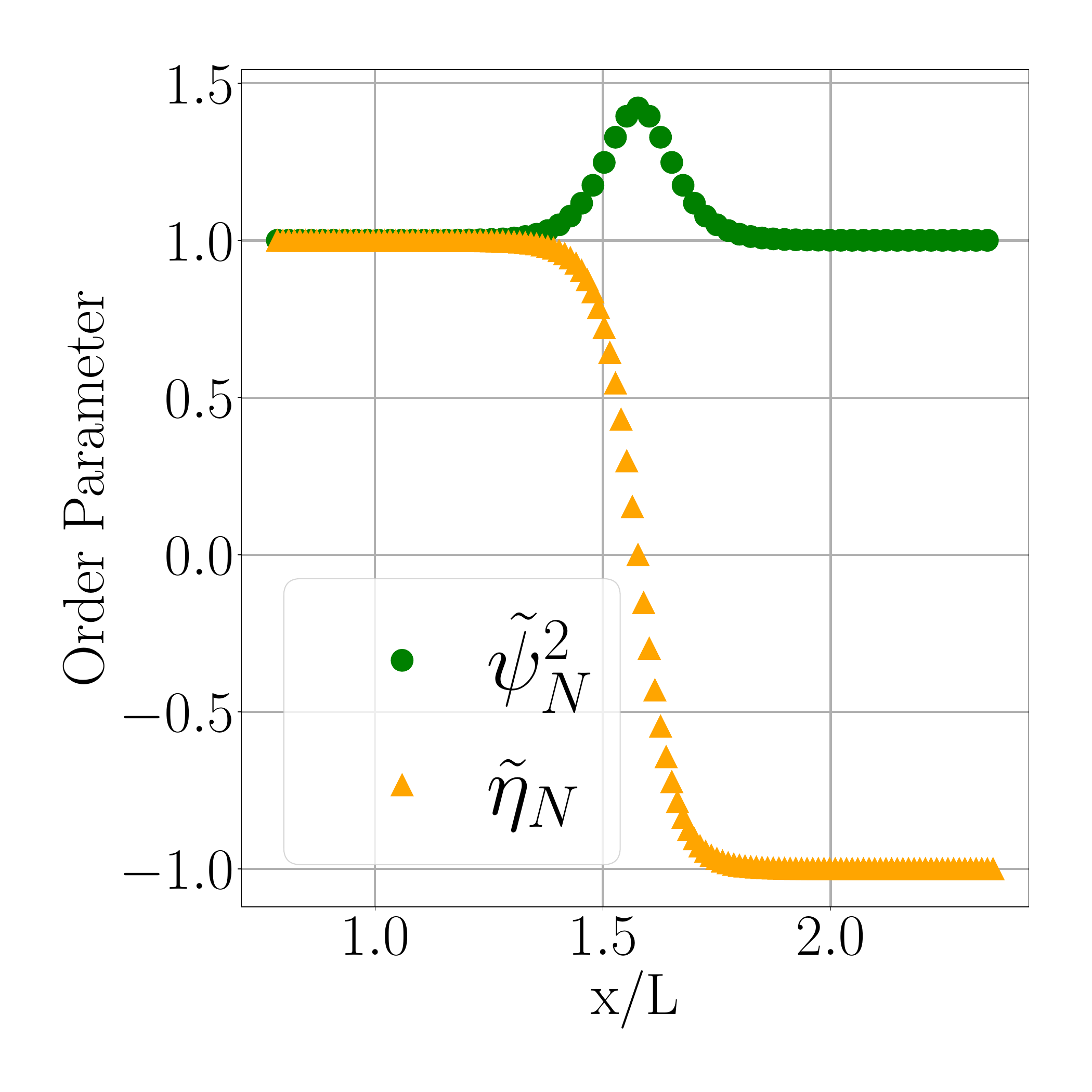}
    }
    \subfloat[\label{fig:fig1c}]{%
      \includegraphics[width=0.48\columnwidth]{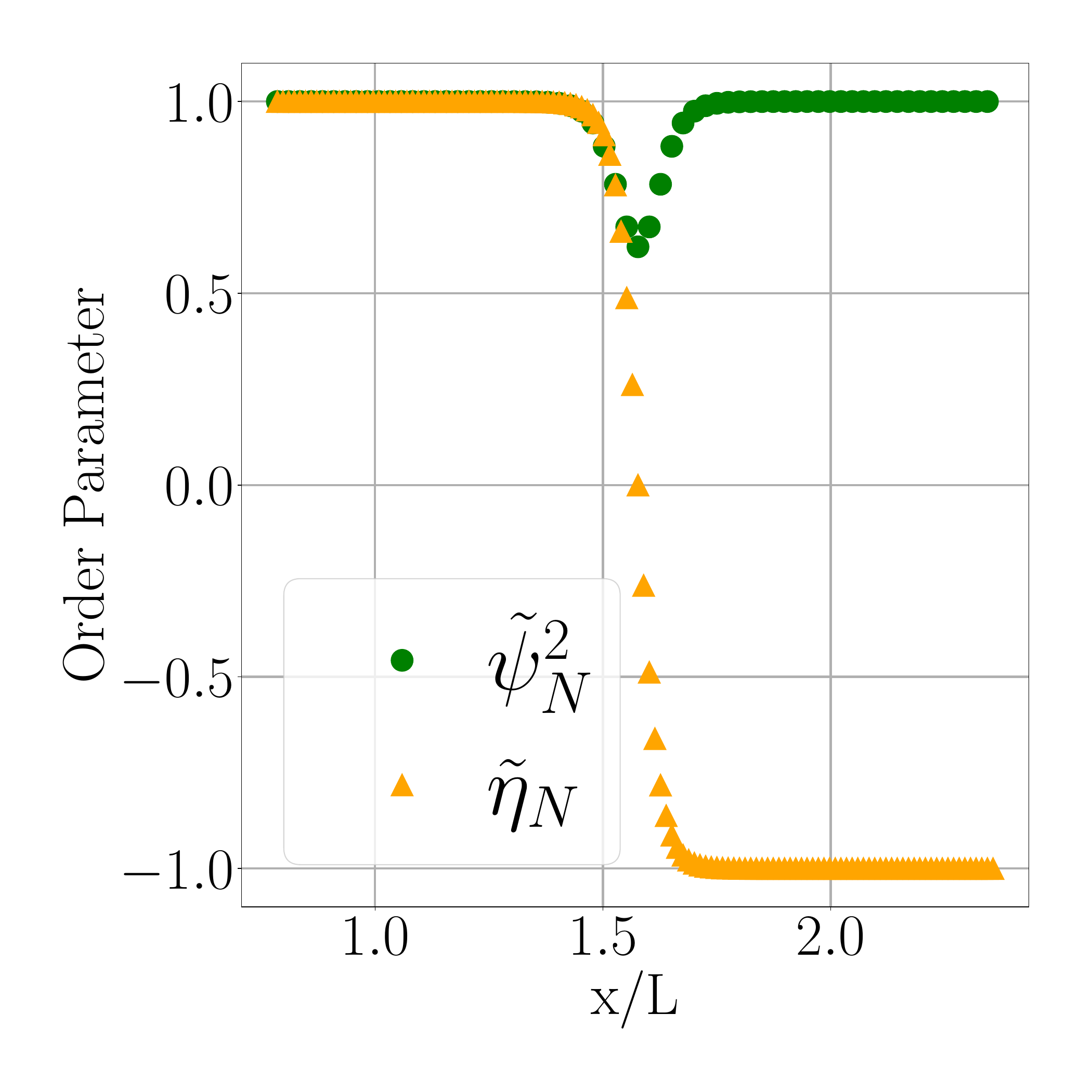}
    }
 \caption{(a) Schematic representation of the domain wall and the broken orthorhombic symmetry. Notice that the crystalline orthorhombic axes at each side of the wall form a 45 degree angle with the wall. (b) Profile of the normalised superconductor $\tilde{\psi}_N^2 = \frac{|\tilde{\psi}|^2}{\tilde{\psi}_v^2}$ and nematic $\tilde{\eta}_N = \frac{\tilde{\eta}}{\Tilde{\eta}_v}$ (see Eqs. (10) and (16) for definitions) order parameters for the competing case with $\hat{\lambda}_2 > 0$ (c) Same as (b) but for the cooperative case with $\hat{\lambda}_2 < 0$ }
   \label{fig:fig1}
\end{figure}
where $F_S$ is the standard energy for an $s$-wave complex superconductor order parameter $\psi$, $F_N$ is the free energy for a real Ising-type nematic order parameter, and $F_{SN}$ is the coupling energy between them. For $F_S$ we write
\begin{equation}
    F_S = \int_V \left[ \alpha_{GL}|\psi|^2 + \frac{\beta}{2}|\psi|^4 + \frac{\hbar^2}{2m}|\mathcal{D}\psi|^2 + \frac{(\nabla\times\mathbf{A})^2}{8\pi} \right] ,
\end{equation}
where the superfluid Cooper pair density is $n_s =|\psi|^2$, $\boldsymbol{A}$ is the magnetic vector potential that relates to the magnetic induction as $\nabla\times\boldsymbol{A} = \boldsymbol{B}$ and $\mathcal{D} = -i\nabla - \frac{e}{\hbar c}\boldsymbol{A}$ is the covariant derivative. Although $m$ stands for a parameter with dimensions of mass, it should not be taken as the mass of any particle or quasiparticle but as a parameter related to the phase stiffness of the order parameter. The charge of the Cooper pairs (twice the electron charge) will be noted as $e$ while $\hbar$ and $c$ are the reduced Planck constant and the speed of light respectively.

We will consider configurations that are translational invariant along the $c$-direction, which implies that any dependence on this coordinate will be hereby ignored. As for the coordinates in the $ab$ plane, attention has to be drawn to the choice of axes. In Fe-based superconductors (as well as in many cuprates), nematicity occurs simultaneously with a tetragonal to orthorhombic structural phase transition, with the $ab$ basis rotated in a $\pi/4$ angle between the different phases. In this work we will choose our $ab$ base to coincide with that one of the tetragonal phase.

The contribution of the nematic Ising type order parameter to the  GL free energy is given by
\begin{equation}
    F_N = \int_V \Big[\gamma_2 (\nabla \eta)^2 + \gamma_3 \eta^2 + \frac{\gamma_4}{2} \eta^4 \Big] ,
\end{equation}
while the coupling between superconductivity and nematicity is given by two terms
\begin{equation}
    F_{SN}=F_{bi}+F_{tri} .
\end{equation}
The first one is a biquadratic coupling of the form 
\begin{equation}
    F_{bi} = \lambda_2\int_V   \eta^2 |\psi|^2 .
 \end{equation}   
 This term indeed does not depend on the  specific character of the nematic order parameter, and it could be present even if $\eta$ was a true scalar field.

 In addition, there could be a trilinear term of the form
  \begin{equation}  
   F_{tr} = \frac{\hbar^2}{2m}\lambda_1 \int_V \eta e_{ij}\mathcal{D}_i\psi  (\mathcal{D}_j \psi)^* ,
\end{equation}
where
\begin{equation}
    e_{ij}=2 (n_i n_j-\frac{1}{2}\delta_{ij}) ,
\end{equation}
and $\mathbf{n}=(\cos\alpha,\sin\alpha)$ is the director signalling orientation of the nematic order with respect to the coordinate basis (which in this work corresponds to the tetragonal phase). A model of this type was used, for instance, in \cite{chowdhury} to study vortices in the London limit, and more recently in \cite{us} where vortex-vortex interactions were analysed.

In the present paper, and inspired by  nematicity in Fe-based superconductors, we will choose  $\alpha$ as follows: As the director aligns according to one of the orthorhombic axes, if we choose $\hat{x}$ and $\hat{y}$ directions to coincide with the tetragonal axis, then $\alpha=\pi/4$ ({\it cf.} \cite{chowdhury,us}, where $\hat{x}$ and $\hat{y}$ were chosen as the orthorhombic axis and then $\alpha=0$). During the phase transition, structural domains with different orientations of the orthorhombic axes can form, separated by a domain wall or twin boundary as shown schematically in Fig \ref{fig:fig1a}.

Note that while the  biquadratic coupling between the nematic and superconductor order parameters can be present even in the case in which $\eta$ is a standard scalar field, the trilinear term is instead a genuine nematic coupling. The biquadratic coupling parameter $\lambda_2$ can be either positive or negative (within some bounds), resulting in competition or cooperation of the nematic order parameter with superconductivity. A non-zero $\lambda_1$ on the other hand  will result in elliptical vortices and its sign will select the orientation of the major axis of the ellipse with respect to the nematic order director. 

\begin{figure*}
    \subfloat{%
      \includegraphics[width=0.31\linewidth]{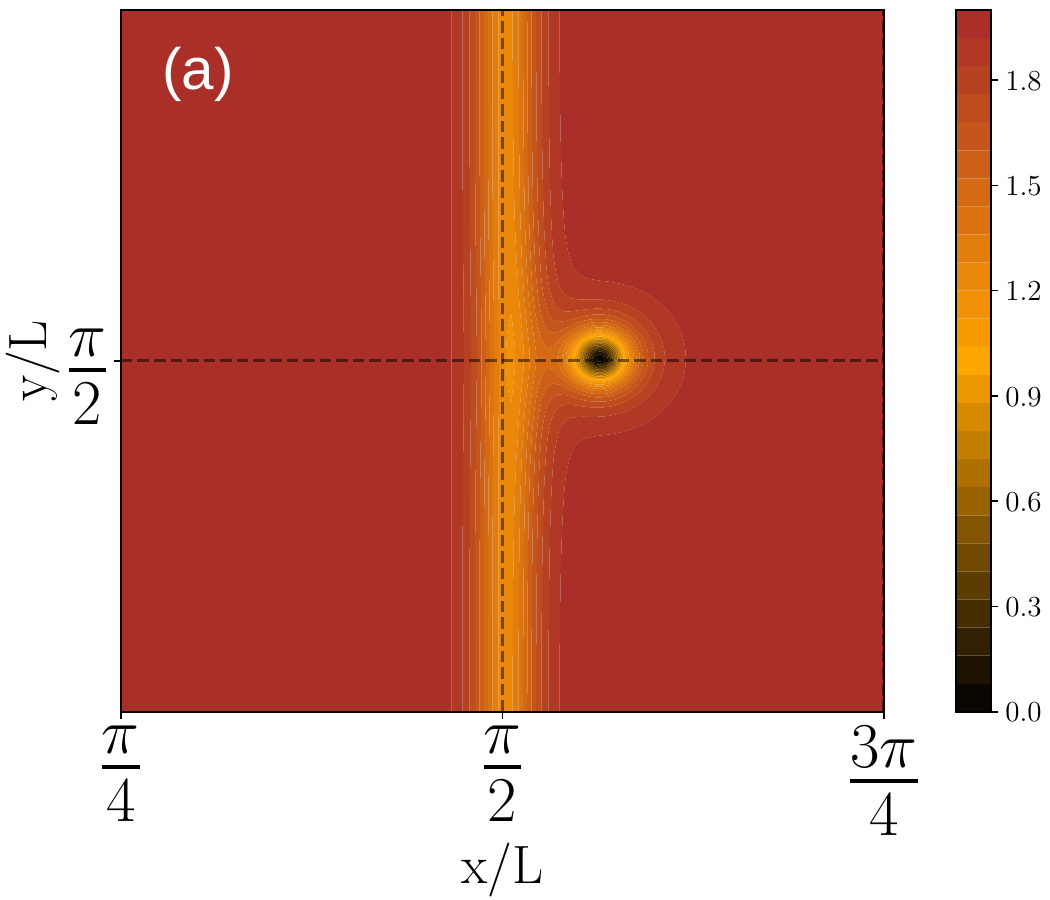}
    }
    \hfill
    \subfloat{%
      \includegraphics[width=0.31\linewidth]{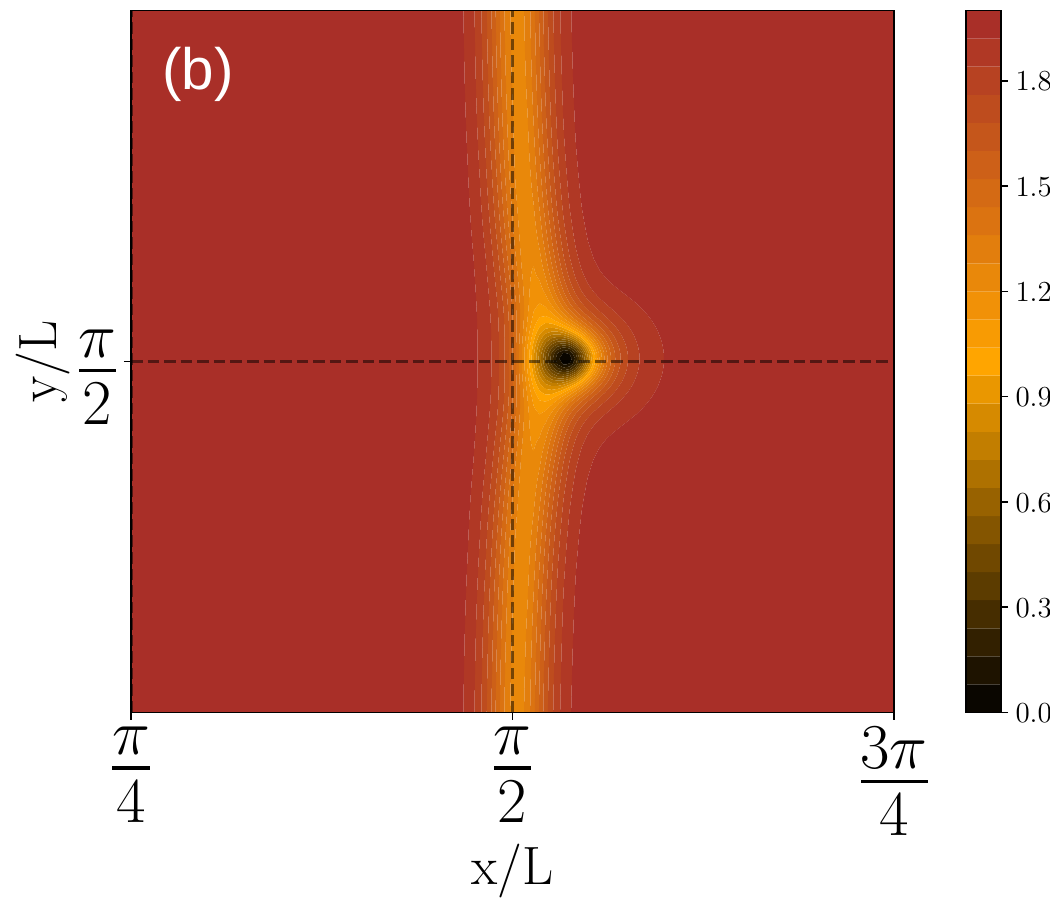}
    }
      \hfill
    \subfloat{%
      \includegraphics[width=0.31\linewidth]{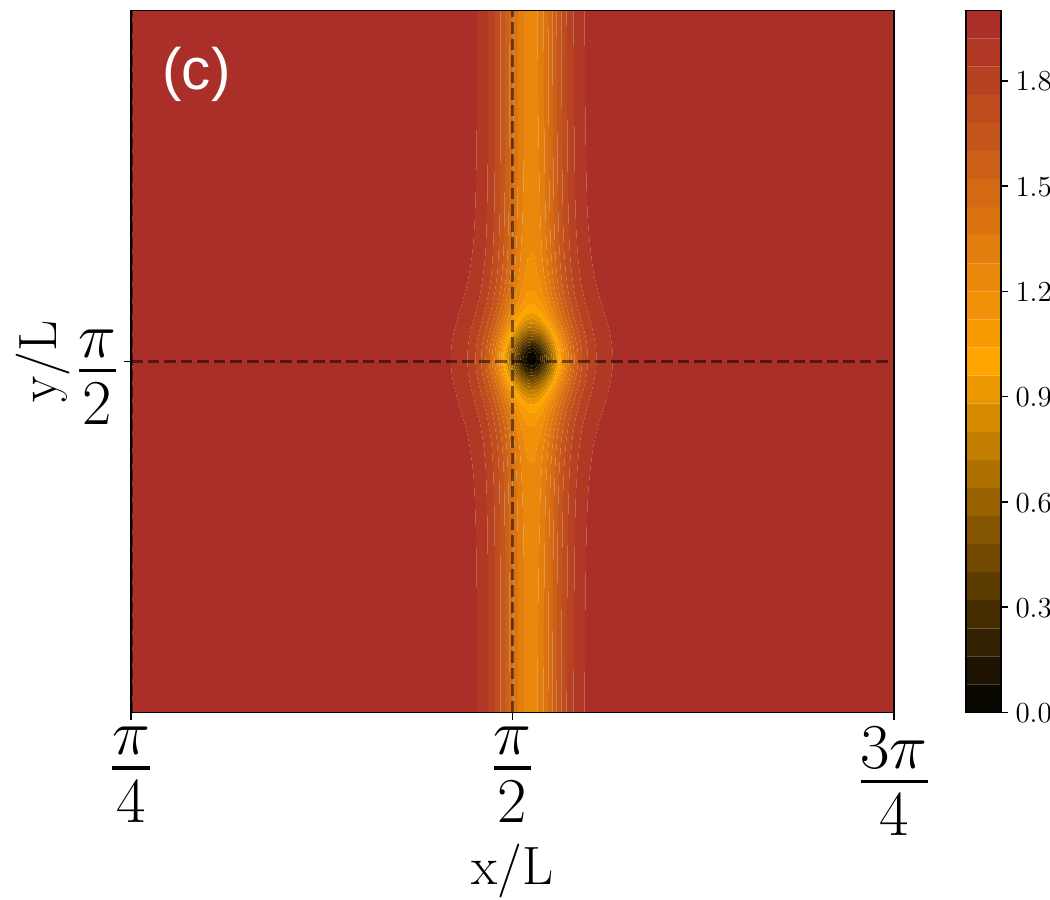}
    }
      \hfill
    \subfloat{%
      \includegraphics[width=0.31\linewidth]{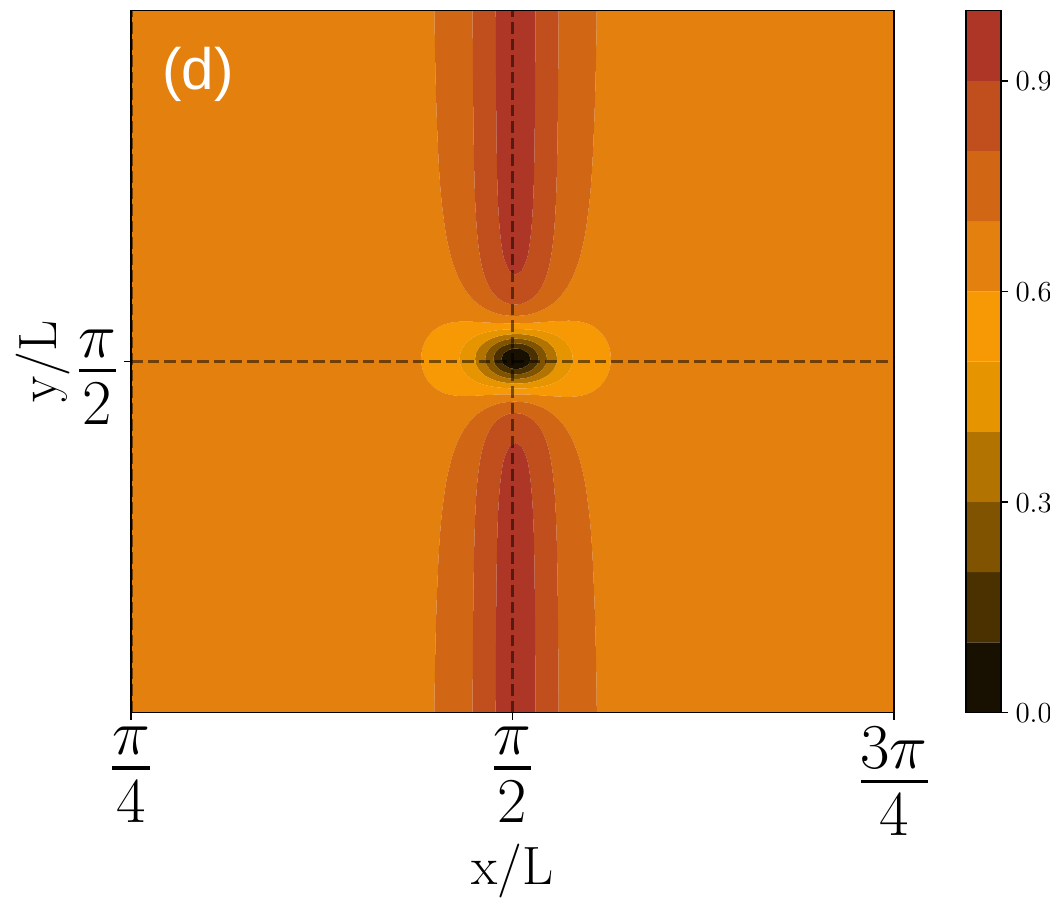}
    }
    \hfill
    \subfloat{%
      \includegraphics[width=0.31\linewidth]{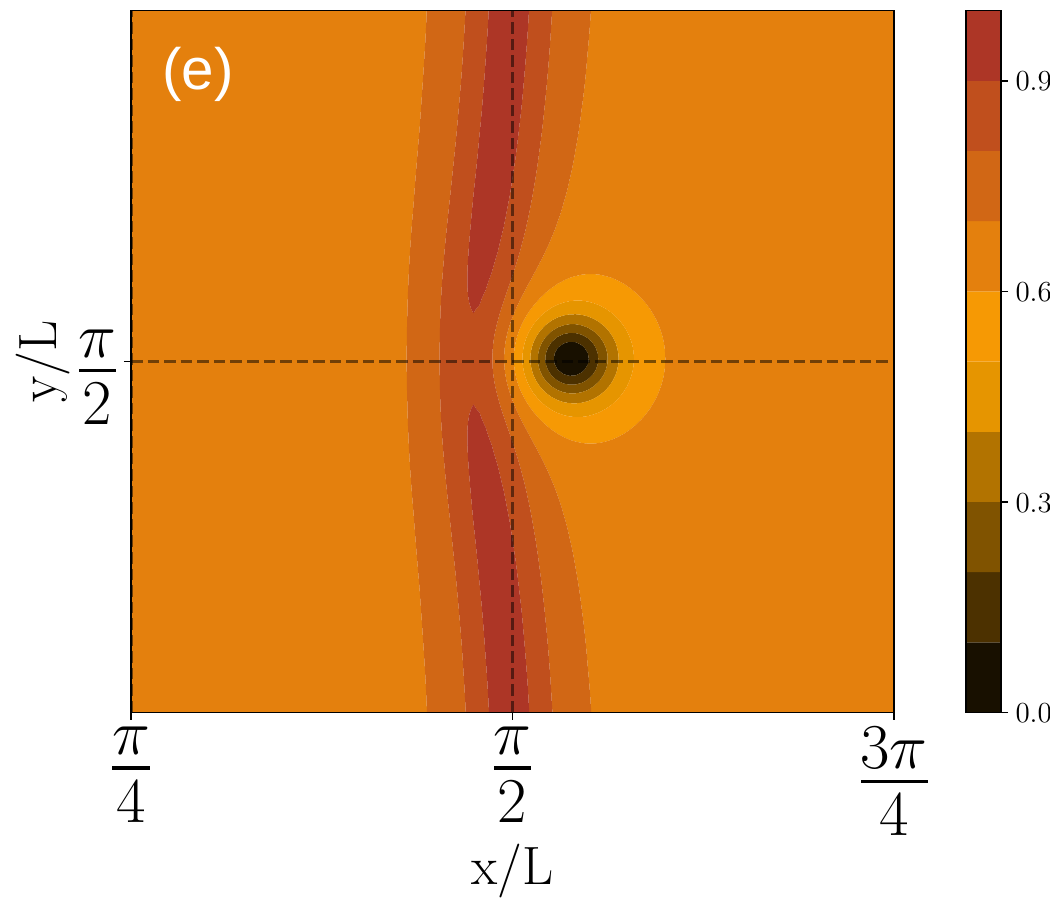}
    }
      \hfill
    \subfloat{%
      \includegraphics[width=0.31\linewidth]{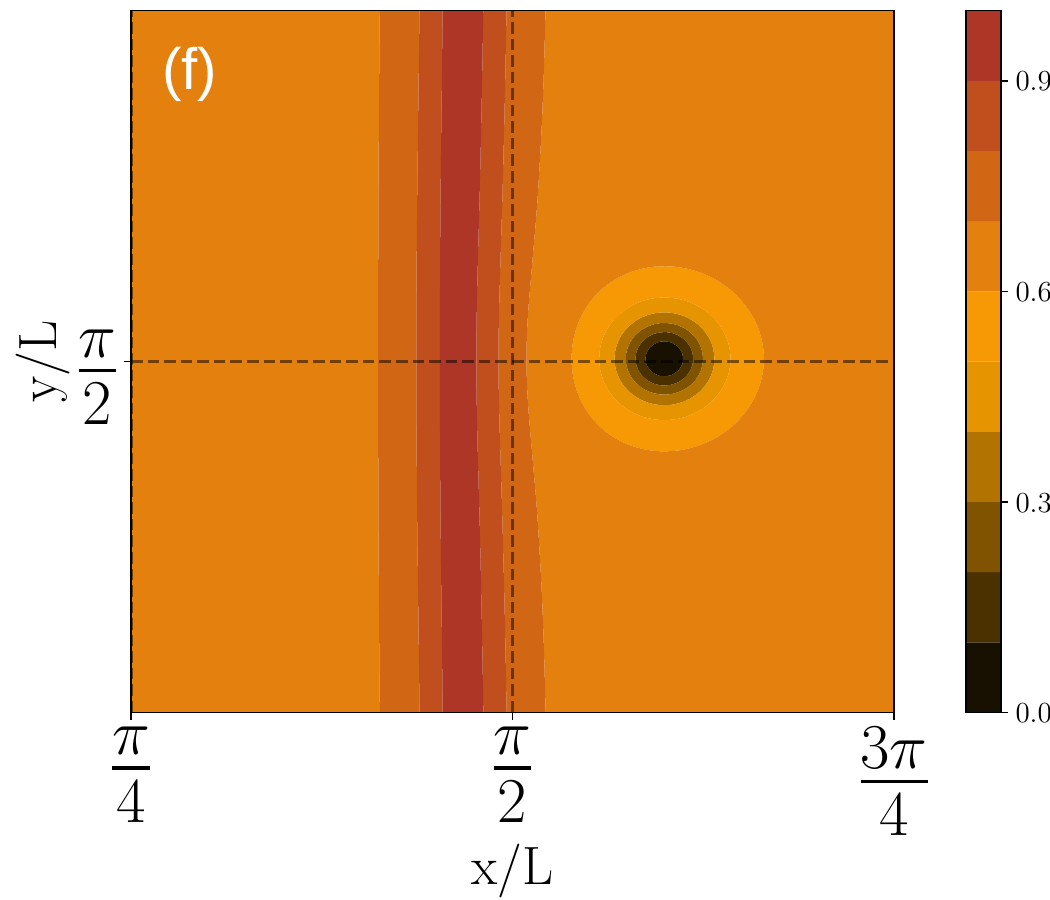}
    }
 \caption{ (\textbf{a, b and c}):   Snapshots of the density plot of $|\tilde{\psi}|^2$ for the purely biquadratic case in the attractive regime with $\hat{\lambda}_2 = -0.5$ and a wall width of $l_\eta = \xi = 0.04$. The vertical (horizontal) dashed line represents the $x = \pi/2$ ( $y = \pi/2$) axis, for visual guidance. Once the vortex is pinned, it changes its symmetry to align with the direction selected by the wall. (\textbf{c, d  and e}): same plots as in \textbf{ a, b and c} but for the repulsive case with $\hat{\lambda}_2 = 0.5$ and the same wall width. The domain wall bends sharply as the vortex is repelled and retakes its original form once the vortex is far from the wall. Notice that once the vortex leaves the domain wall it retains its cylindrical symmetry. (see text below).}
   \label{fig:bic_rep_att}
  \end{figure*}
  
As we did in \cite{us}, we consider dissipative dynamics \cite{schmid} and write the time dependent equations for the order parameters and the electromagnetic fields as
\begin{equation}
    \frac{\hbar^2}{2mD}\partial_t \psi = -\frac{\delta F}{\delta \psi^{*}} , \,
    \frac{\sigma}{c^2} \partial_t \boldsymbol{A} = -\frac{\delta F}{\delta \boldsymbol{A}} , \,
    \frac{\hbar^2}{2m D_n} \partial_t \eta = - \frac{\delta F}{\delta \eta} .
\end{equation}
In the above expressions, $D$ and $D_n$ are two diffusion constants (one for each order parameter, and not necessarily equal), and $\sigma$ is the electrical conductivity of the normal state.  

Three different length scales are present in the system, namely
\begin{equation}
    \xi^2 = \frac{\hbar^2}{2m|\alpha_{GL}|}, \quad \quad \lambda_L^2 = \frac{mc^2}{4\pi e^2 \rho_0}, \quad \quad l_\eta^2 = \frac{\gamma_2}{|\gamma_3|} ,
    \label{length}
\end{equation}
where $\xi$, $\lambda_l$, and $l_\eta$ are the bare superconductor coherence length, the bare London penetration length (where $\rho_0 = \frac{|\alpha_{GL}|}{\beta_{GL}}$), and the bare nematic coherence length respectively. We can rewrite the superconductor and nematic order parameters in terms of two dimensionless order parameters $\tilde{\psi}$ and $\tilde{\eta}$ as
\begin{equation}
    \psi = \sqrt{\rho_0} \tilde{\psi}, \quad \quad \eta = \eta_0 \tilde{\eta},
\end{equation}
where $\eta_0^2 = \frac{|\gamma_3|}{\gamma_4}$, and redefine the vector potential in terms of a simpler quantity $\mathbf{a}$  as
\begin{equation}
    \boldsymbol{A} = \frac{mc|\alpha_{GL}|}{\hbar e}\mathbf{a} .
\end{equation}

In terms of the newly defined variables, the total free energy can be written as:
\begin{equation}
\begin{split}
     & F = |\alpha_{GL}| \rho_0 \int_V  \frac{1}{2}(|\tilde{\psi}|^2-1)^2+\xi^2 |\nabla\tilde{\psi}|^2 - \boldsymbol{a}\text{Im}(\tilde{\psi}^{*}\nabla\tilde{\psi})+ \\ 
     & \frac{1}{4\xi^2}\boldsymbol{a}^2\tilde{\psi}^2+\frac{\kappa^2}{4} (\nabla\times\boldsymbol{a})^2+\Gamma_2(\nabla\tilde{\eta})^2+\frac{\Gamma_4}{2}(\tilde{\eta}^2-1)^2+\\
     & \hat{\lambda}_2 \tilde{\eta}^2\tilde{\psi}^2-\xi^2\hat{\lambda}_1\tilde{\eta}[(\mathcal{D}_x\psi)^{*}\mathcal{D}_y\psi + (\mathcal{D}_y\tilde{\psi})^{*}\mathcal{D}_x\tilde{\psi} ] ,
\end{split}
\end{equation}
where as we mentioned before,  we have set the director angle for the nematic parameter as $\alpha=\pi/4$, 
and have redefined parameters as
\begin{equation}
\begin{split}
     &\Gamma_2 = \frac{\gamma_2}{|\alpha_{GL}|(\rho_0/\eta_0^2)}, \quad \Gamma_4 = \frac{\gamma_4 \eta_0^2}{|\alpha_{GL}|(\rho_0/\eta_0^2)}, \\
     &\hat{\lambda}_2 = \frac{\lambda_2 \eta_0^2}{|\alpha_{GL}|},  \qquad \hat{\lambda}_1 = \frac{\lambda_1\eta_0}{\hbar}. 
\end{split}
\end{equation}
We have also introduced the GL parameter in its usual definition as
\begin{equation}
    \kappa = \frac{\lambda_L}{\xi}.
\end{equation}
Notice that these redefinitions imply that $\Gamma_2$ has dimensions of $[L]^2$, while the rest of the parameters are dimensionless. The nematic coherence length with these new variables is calculated as $l_\eta^2 = \frac{\Gamma_2}{\Gamma_4}$.

We will work under the assumption that both superconducting and nematic symmetries are broken. For constant order parameters, the minimum energy is achieved when the interaction potential is minimised, 
\begin{equation}
    \begin{split}
        V(|\tilde{\psi}|^2, \tilde{\eta}) =& \rho_0 |\alpha_{GL}| \int_V \frac{1}{2}(|\tilde{\psi}|^2 - 1)^2+\\
        &\frac{\Gamma_4}{2}(\tilde{\eta}^2 -1)^2  +\hat{\lambda}_2 |\tilde{\psi}|^2 \tilde{\eta}^2 .
    \end{split}
\end{equation}
On the minimum, the order parameters take the values
\begin{equation}
    \tilde{\psi}_v^2=\frac{1-\hat{\lambda}_2}
    {1-\frac{\hat{\lambda}_2^2}{\Gamma4}}, \quad
    \tilde{\eta}_v^2=\frac{1-\frac{\hat{\lambda}_2}{\Gamma_4}}
    {1-\frac{\hat{\lambda}_2^2}{\Gamma4}},
\end{equation}
and thus the free energy is
\begin{equation}
    F_0=|\alpha_{GL}| \rho_0 V \frac{(\tilde{\psi}_v^2-1)(\tilde{\psi}_v^2+\tilde{\eta}_v^2)}{2 \tilde{\eta}_v^2}.
\end{equation}
For further convenience we will refer the energy to this value,
\begin{equation}
    \tilde{F}= F-F_0 .
\end{equation}
The theory can be reformulated in terms of a dimensionless time parameter $\tau$, related to the physical time $t$ by
\begin{equation}
    t = \frac{\hbar^2}{2mD|\alpha_{GL}|} \tau ,
\end{equation}
if we also redefine the conductivity and the nematic difussion constant as
\begin{equation}
    \sigma_1 = \frac{4\pi \sigma}{c^2} \frac{2mD|\alpha_{GL}|}{\hbar^2} , \quad \quad D_\eta = D_n (\rho_0/\eta_0^2).
\end{equation}
The explicit form of the equations can be found in \cite{us}.

\section{ Vortex-domain wall interaction}\label{vortex}

\subsection{Numerical method}\label{numerical}

Calculations were performed using the Geophysical High-Order Suite for Turbulence (or GHOST for short \cite{ghosta,ghostb}), which uses a high-order pseudo-spectral method. Following Ref.~\cite{nore} we consider configurations in a  $[0,2\pi] \times [0,2\pi]$ simulation box using $512 \times 512$ grid points. We only focus on a $[0,\pi]\times [0,\pi]$ sector, since the other sectors serve as images to satisfy the periodicity in the $[0,2\pi] \times [0,2\pi]$ domain required by the pseudo-spectral method. The initial conditions depend on the number of singularities (domain walls or superconducting vortices) under consideration. 

As discussed in \cite{us}, the initial condition for a single vortex located at $(x/L, y/L) = (\pi/2, \pi/2)$ is taken as 
\begin{equation}
    \tilde{\psi}(x, y, t = 0) = \tilde{\psi}_v \frac{(\lambda + i \mu)}{\sqrt{\lambda^2 + \mu^2}} \tanh{\frac{\sqrt{\lambda^2 + \mu^2}}{\sqrt{2}\xi}},
\end{equation}
where the Clebsch potentials are $\lambda(x) = \sqrt{2}\cos{x}$ and $\mu(y) = \sqrt{2}\cos{y}$ (a shift in the vortex position is trivially achieved by modifying the Clebsch potentials), and  the initial condition for the vector potential is
\begin{eqnarray}
    a_x (x, y, t=0) = a_0 \sin{(x)}\cos{(y)} \label{eq:initialax}, \\
    a_y (x, y, t=0) = - a_0 \cos{(x)}\sin{(y)}\label{eq:initialay},
\end{eqnarray}
which is a well known initial condition in fluid dynamics known as a Taylor-Green flow. The resulting initial magnetic field $B_z(x,y,t=0) = 2 a_0 \sin{x}\sin{y}$ (with $a_0$ a normalisation constant related to the total flux in the simulation box, see \cite{us}) satisfies the necessary conditions of periodicity and is easy to implement numerically.

For the simulations carried out in this paper we set the nematic coherence length to be equal to the superconducting coherence length $l_\eta = \xi = 0.04$. We set $\kappa = \frac{4}{\sqrt{2}}$ (this choice might no be realistic for real superconductors, but makes simulations simpler as all relevant length scales are of the same order).
For simplicity we have also set  $\rho_0=\eta_0=1$ and $\Gamma_4 = 1$ (notice that this choice also sets the value of $\Gamma_2$). On the other hand, we have chosen  $D_n = D = \frac{\hbar}{2m}$, and the dimensionless conductivity as $\sigma_1 = 15$. The values of these constants can be relevant in setting time scales for equilibration processes and energy dissipation but will not alter the main conclusions of this work. An integration time step  $dt = 6\times 10^{-5}$ was used, consistent with the required Courant–Friedrichs–Lewy condition for stability of the numerical solution (see \cite{us}).

\subsection{The superconductor order parameter and the domain wall}\label{the superconductor}

\begin{figure*}
    \subfloat[\label{fig:fig3a}]{%
      \includegraphics[width=0.32\linewidth]{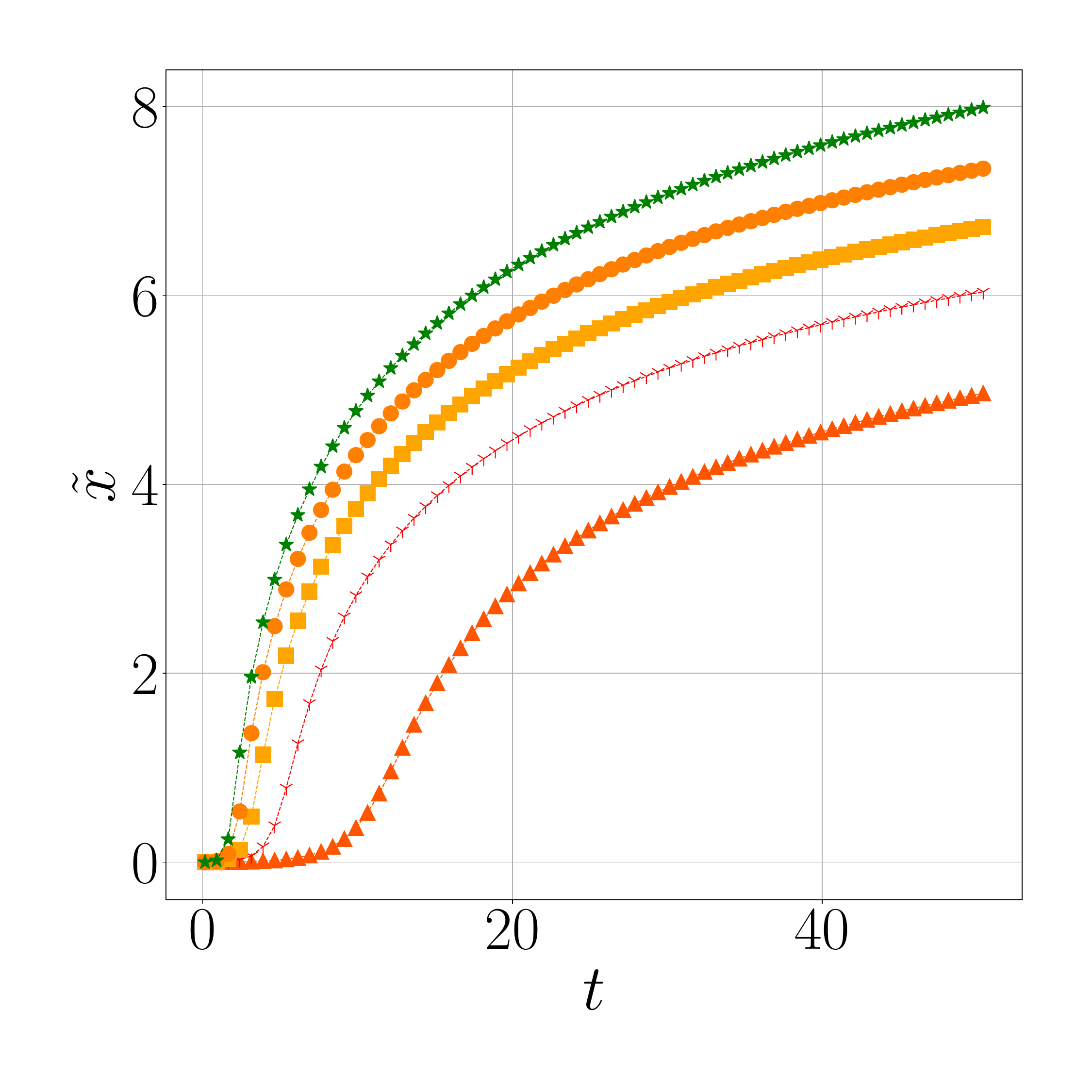}
    }
    \hfill
    \subfloat[\label{fig:fig3b}]{%
      \includegraphics[width=0.32\linewidth]{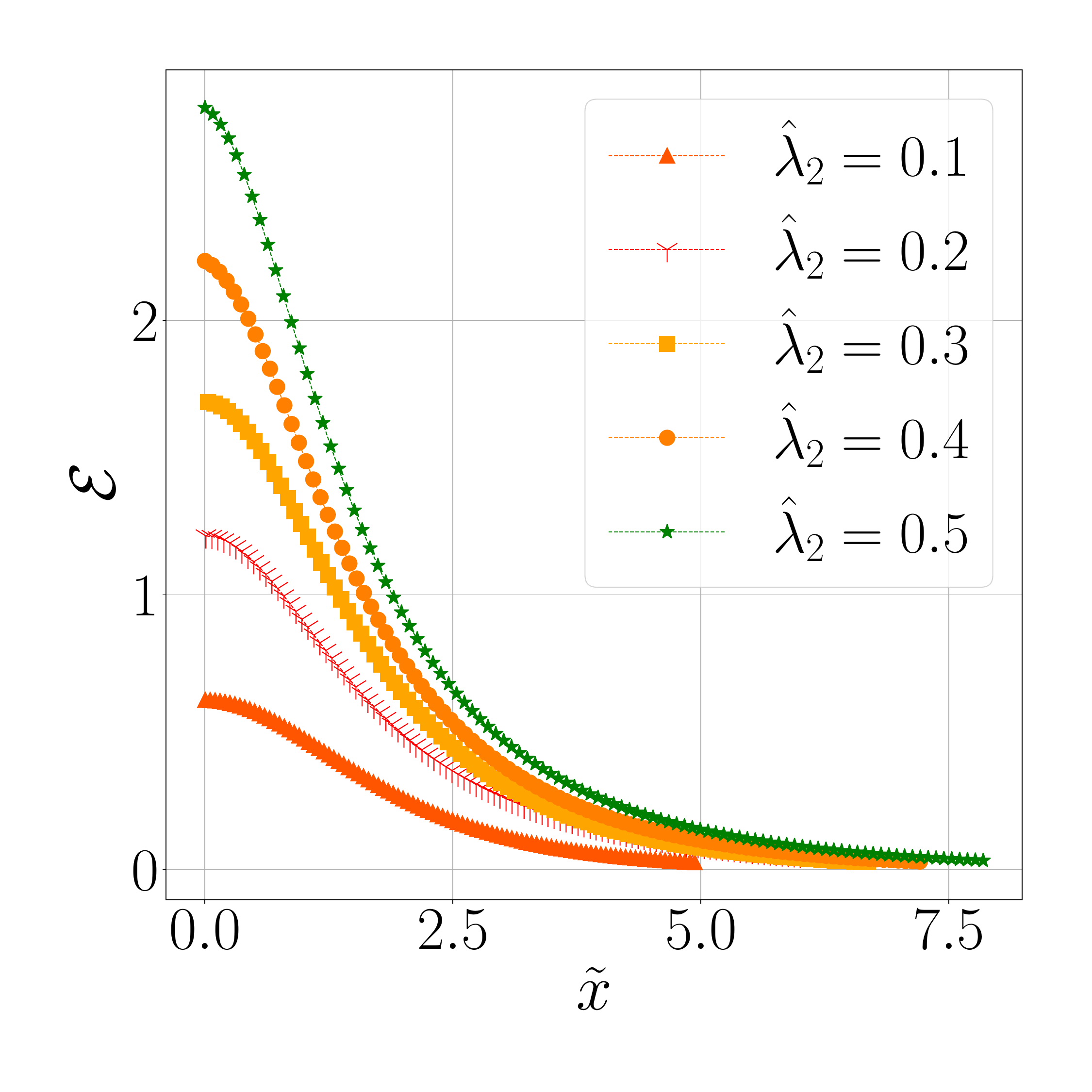}
    }
    \hfill
    \subfloat[\label{fig:fig3c}]{%
      \includegraphics[width=0.32\linewidth]{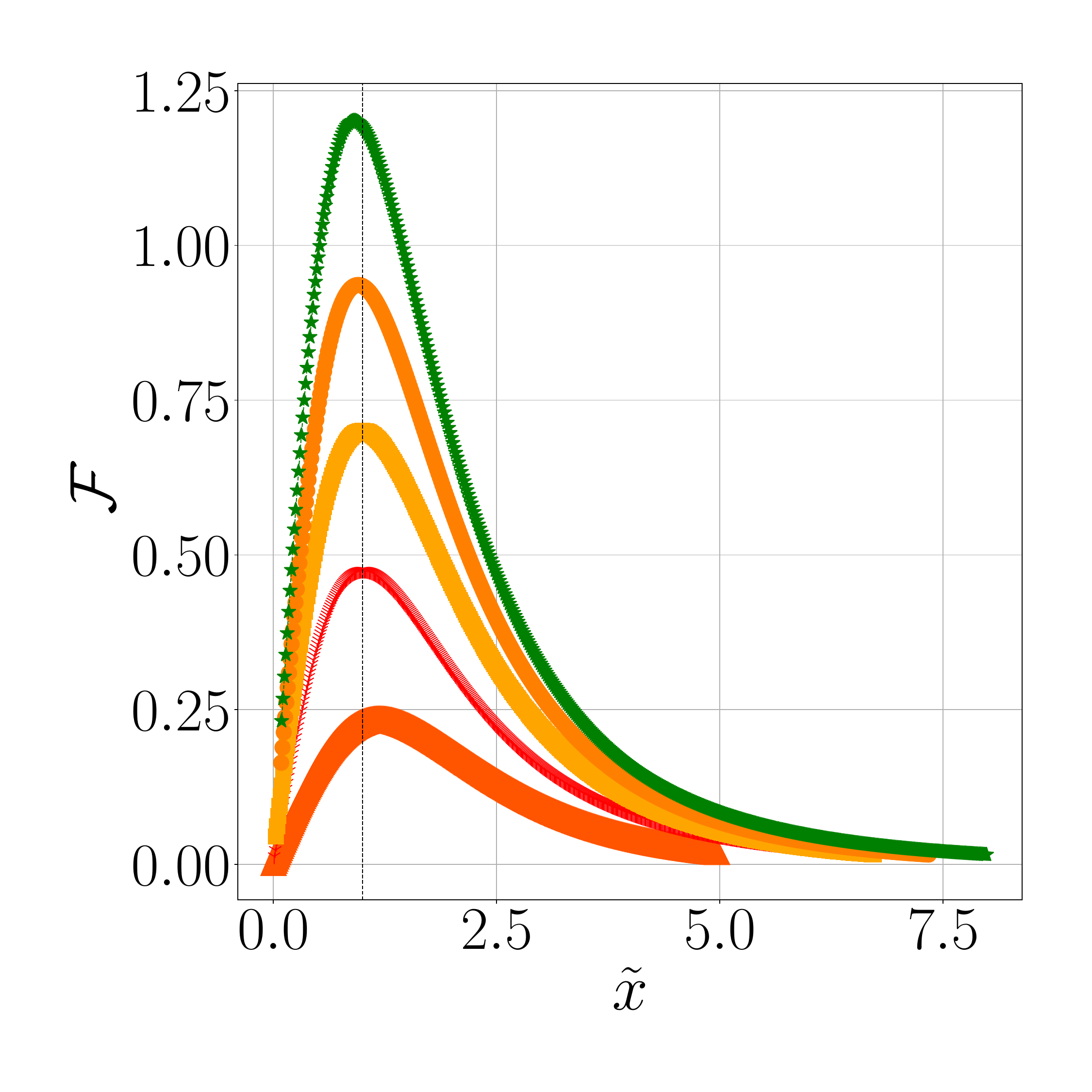}
    }
 \caption{\textbf{(a)} Distance of the vortex from the NDW, measured from the center of the wall, as a function of time for different values of $\hat{\lambda}_2$ (with $\hat{\lambda}_1 = 0$) \textbf{(b)} Dimensionless interaction energy as a function of the distance of the vortex to the center of the NDW (see Eqs.~(27-30) for definitions) \textbf{(c)} Dimensionless interaction force as a function of the vortex distance from the NDW. The dashed vertical line indicates the domain wall edge.}
     \label{fig:bic_analisis}
  \end{figure*}

It is well known that an Ising type order parameter theory admits non trivial solutions representing a two dimensional defect, or a {\em domain wall}. For instance,  it is easy to check that when $\hat{\lambda}_1 = \hat{\lambda}_2 = 0$ a static solution for $\tilde{\eta}$ is
\begin{equation}
   \tilde{ \eta}(x) = \tilde{\eta}_v \tanh{\left(\frac{x}{\sqrt{2}l_\eta}\right)},
    \label{eq:domain_exact}
\end{equation}
which represents a domain wall located at $x=0$, following the typical orientation of a structural twin boundary (the position and orientation  of the wall can be easily changed). Here,  $l_\eta$ is the nematic coherence length defined in Eq.~(\ref{length}) and related to the domain wall thickness. 

In this subsection we explore the interaction of a domain wall of this type with a superconducting vortex, placed at different distances from the nematic domain wall. In order to induce the domain wall formation at the $x/L = \pi/2$ plane,  we take the initial condition corresponding to the nematic order parameter as 
\begin{equation}
   \tilde{ \eta}(x,y, t=0) = \frac{\tilde{\eta}_v}{\tanh{\left(\frac{1}{\sqrt{2}l_\eta}\right)}} \tanh{\left(\frac{\lambda(x)}{2 l_\eta}\right)} .
    \label{eq:eta_ini}
\end{equation}

As a first test of the numerical method, we verified that for the pure nematic theory the initial condition  (\ref{eq:eta_ini}) evolves to the exact solution
(\ref{eq:domain_exact}). The energy per area of the domain wall can be calculated explicitly, resulting
 \begin{equation}
    E_{wall} = \frac{4}{3}\rho_0 |\alpha_{GL}|\sqrt{2\Gamma_2 \Gamma_4} ,
    \label{eq:wall_energy}
 \end{equation}
where the area of the wall is $L_y \times L_z =\pi^2$. Our simulations indicate that the initial condition proposed for the nematic order parameter rapidly converges to the exact solution for the wall, and the energy can be numerically computed with precision down to $\mathcal{O}(10^{-4})$.

We continue studying the behaviour of the  superconductor order parameter in the presence of a domain wall. We first consider the case in which $\lambda_2 \neq 0$, $\lambda_1=0 $, and the initial configuration for the superconductor order parameter is $|\tilde{\psi}|^2 = \rho_0$ (i.e, there is no vortex). We show in Fig.~\ref{fig:fig1b} (Fig.~\ref{fig:fig1c}) that  superconductivity is enhanced (depressed) for positive (negative) values of $\hat{\lambda}_2$. This is consistent with the fact that the order parameters compete (cooperate) for positive (negative) values of $\lambda_2$.

\subsection{A single vortex in the presence of a nematic domain wall}\label{single vortex}

We now turn to study the case of a single vortex in the presence of a nematic domain wall. We first study the effect of the biquadratic coupling and after that we see how these results are modified when the  trilinear coupling is active. As we mentioned before, the biquadratic coupling would be present even in cases where the order parameter was a scalar, so results in this subsection could apply to a more general range of theories.

\subsubsection{Biquadratic coupling}\label{biquadratic}

Before entering into the details of the simulations, it is clear that for $\lambda_2 >0$  ($\lambda_2 <0$) the interaction will be repulsive (attractive). Because superconductivity is enhanced for positive coupling, the vortex will move its normal core away from the wall in order to minimise the free energy, thus exhibiting a repulsive interaction between the vortex and the domain wall. On the other hand, since superconductivity is depressed for negative coupling, we expect that the normal core will tend to remain on the domain wall, resulting instead in an  attractive interaction. It is also easy to predict that the interaction is short ranged.
 
We show in the top panel of Fig.~\ref{fig:bic_rep_att} the density plot of $|\tilde{\psi} (x,y)|^2$ for the attractive case ($\lambda_2 <0$). We start with a vortex at a distance $d = 3\xi$ from the domain wall, and after a finite time the vortex becomes pinned to the wall. Notice that the wall has its own dynamics, and it bends during the pinning process. The rigidity of the wall is controlled by the parameters of the theory. In the bottom panel of Fig.~\ref{fig:bic_rep_att} we show the same density plot but for the repulsive case ($\lambda_2>0$). For this simulation we placed the vortex within the domain wall and waited until it was repelled. Remember that the effective superconducting coherence length depends on the value and sign of $\lambda_2$, and thus the core size is noticeable larger than in the attractive case. The wall bending effect is also more pronounced in the repulsive than in the attractive case. 
 
Far from the domain wall the vortex core has cylindrical symmetry, but as it is attracted to the NDW the vortex loses this symmetry and ends up  elongated along the direction selected by the domain wall. The lack of cylindrical symmetry is more evident when the $\lambda_1$ coupling is turned on, as we will discuss soon. Since there is no direct coupling between the vector potential and the nematic order parameter, the lack of cylindrical symmetry is less pronounced in the magnetic field (not shown). 

\begin{table}
\begin{ruledtabular}
    \begin{tabular}{lccc}
$\hat{\lambda}_2$ & {$\mathcal{E}_0$} & $c_1$ & $c_2$ \\
\hline
0.1 & 0.617 & 1.658 & 2.434 \\
0.2 & 1.176 & 2.490 & 2.757 \\
0.3 & 1.703 & 3.358 & 3.044  \\
0.4 & 2.223 & 4.546 & 3.413 \\
0.5 & 2.761 & 6.328 & 3.922 \\
\end{tabular}
\end{ruledtabular}
\caption{Fitting parameters for the energy as a function of the vortex-domain wall distance in the biquadratic repulsive regime (see text for details).}
\label{table:fitting_table}
\end{table}

We now characterise the interaction between the vortex and the domain wall in a more quantitative way. For the sake of clarity we mostly focus on the repulsive case (a similar analysis can be performed for the attractive case). In Fig.~\ref{fig:bic_analisis} we show the trajectory of the vortex core as a function of time, the interaction energy per unit length, and the force per unit length as a function of the distance of the vortex core to the wall, for different values of the biquadratic coupling parameter.
For these simulations the vortex was placed on top of the domain wall in an unstable equilibrium, signed by the transient time that it takes for the vortex to begin its movement. We see that the vortex acquires maximum acceleration at the boundary of the wall, as shown in Fig.~\ref{fig:fig3a}.
  
To calculate the interaction energy we first calculate the domain wall energy and the energy of an isolated vortex (with no domain wall), and we then substract these energies from  the total energy in the simulation domain. More precisely, we are interested in energies per unit length of the vortex. As we are assuming traslational symmetry along the $c$ axes, this amounts to integrating over the $ab$ plane. Notice that we defined dimensionless coordinates $\tilde{x}=x/\xi$, $\tilde{y}=y/\xi$, and field $\tilde{\mathbf{a}}=\mathbf{a}/\xi$.
Thus, energies per unit length can be expressed as 
\begin{equation}
    \tilde{E}={\tilde{E}_0} \mathcal{E},
\end{equation}
where $\mathcal{E}$ is a dimensionless function of the variables $\kappa$, $\hat\lambda_1$, $\hat\lambda_2,\Gamma_4 $, and $\Gamma_2/\xi^2$, and $\tilde{E}_0$ sets the scale for the energy per unit length, 
\begin{equation}
\tilde{E}_0=|\alpha_{GL}| \rho_0 \xi^2=\frac{|\alpha_{GL}|^2}{\beta} \xi^2=\frac{H_c^2}{4 \pi} \xi^2,
\end{equation}
where $H_c$ is the thermodynamic superconducting critical magnetic field. 

We can define the dimensionless interaction energy between the vortex and the NDW as:
\begin{equation}
    \mathcal{E}_{int} = \frac{\tilde{E}_T - \tilde{E}_{vortex} - \tilde{E}_{wall}}{\tilde{E_0}} .
\end{equation}
Using this data we can reconstruct $\mathcal{E}_{int}(x)$ and calculate the effective force per unit length $\tilde{F}_{eff}$ between the vortex and the wall as
 \begin{equation}
     \tilde{F_{eff}}=-\frac{\partial \mathcal{E}_{int}}{\partial x}=F_0 \mathcal{F}=\frac{H_c^2}{4 \pi} \xi \mathcal{F},
 \end{equation}
where $ \mathcal{F}$ is a dimensionless function.
The data can be fitted by a simple expression of the type
\begin{equation}
\mathcal{E}_{fit}=\frac{\mathcal{E}_0}{1+c_1 
   {\sinh^2{\frac{x}{c_2}}}}.
\end{equation}
The coefficients that fit the aforementioned energies can be found in Table \ref{table:fitting_table}.

Figure \ref{fig:fig3c} shows the calculated interaction force between the domain wall and the vortex, where we have scaled the position in units of the coherence length. Notice that the maximum of the force occurs in the vicinity of the domain wall edge, marked by the dashed vertical line in Fig.~\ref{fig:fig3c}. The force exponentially tends to zero as the vortex leaves the wall, confirming the short range nature of the interaction.

\subsubsection{The effect of the trilinear term}\label{trilinear}

We next study the case in which  the $C_4$ breaking coupling is present. It has been observed in many experiments (see below) that vortices are elliptical with the axes of the ellipsoid oriented along the axes of the orthorhombic phase. The trilinear term in our model reproduces that effect. To see this we take the trilinear term with $\eta$ constant and fixed. Then, the free energy is
\begin{figure*}[!ht]
    \subfloat{%
      \includegraphics[width=0.32\linewidth]{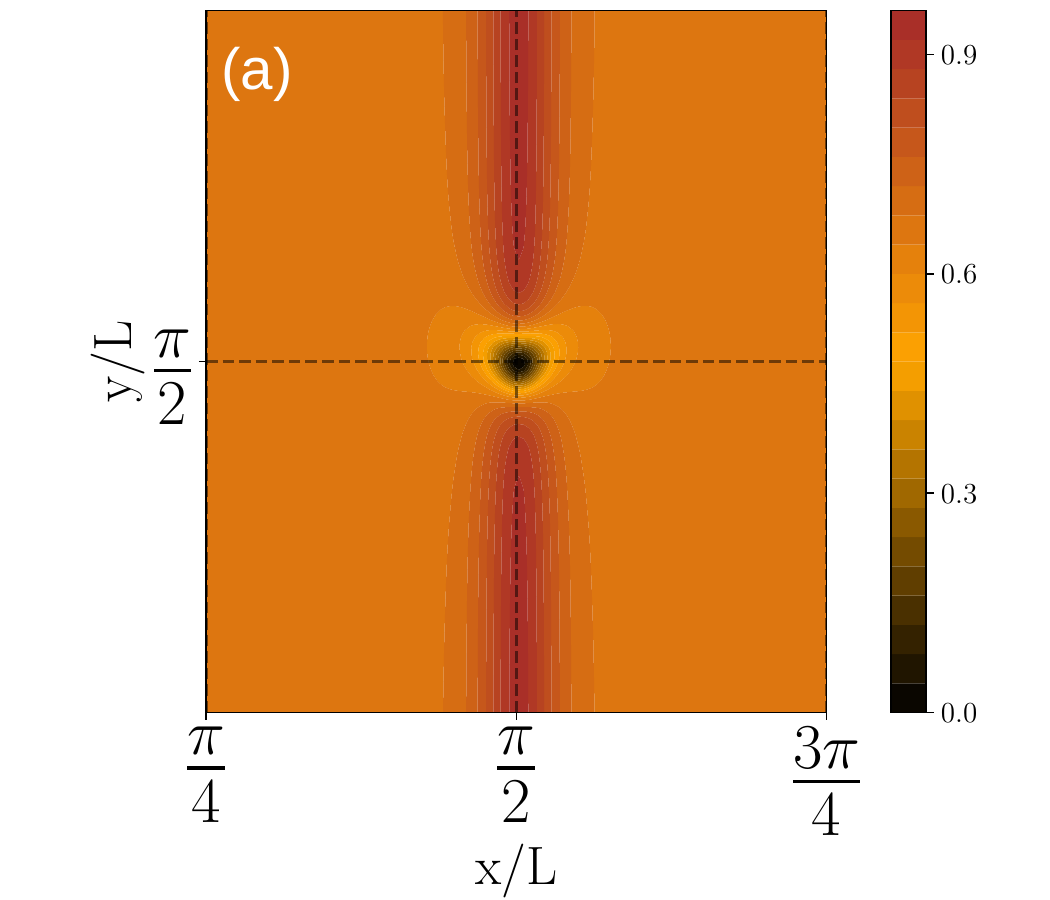}
    }
    \subfloat{%
      \includegraphics[width=0.32\linewidth]{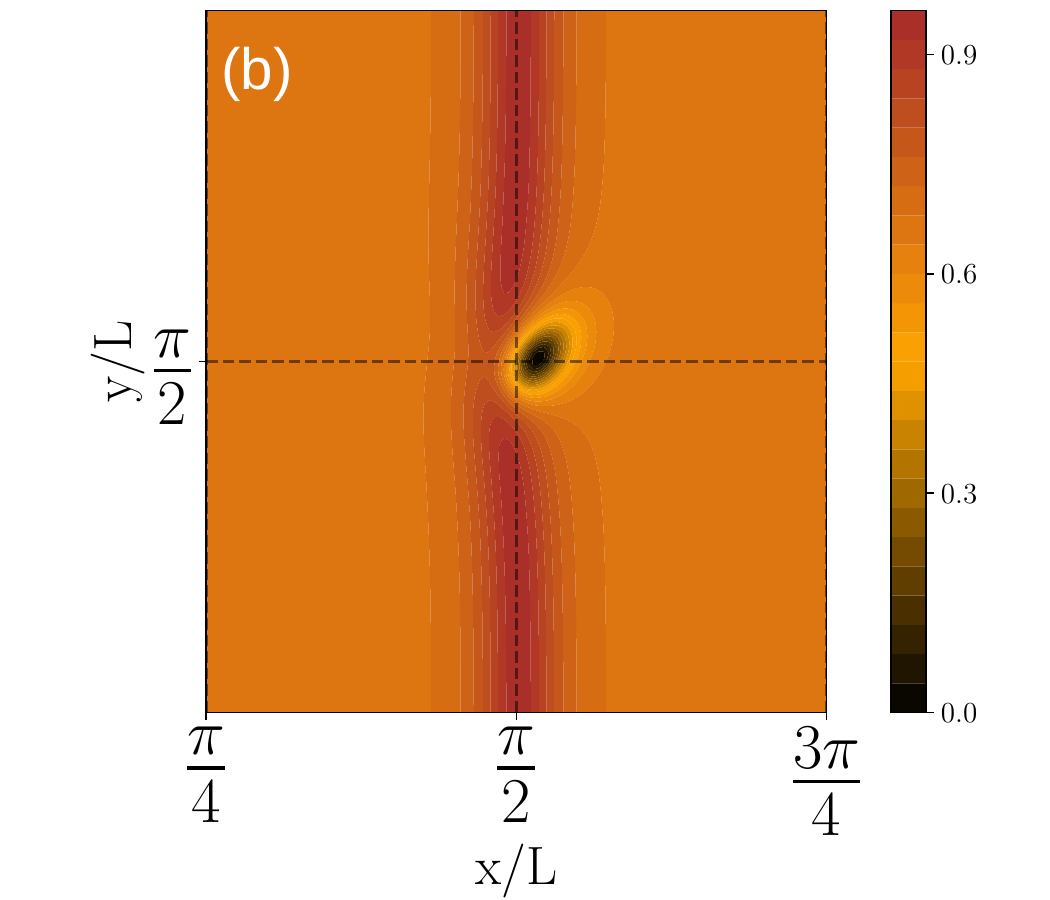}
    }
    \subfloat{%
      \includegraphics[width=0.32\linewidth]{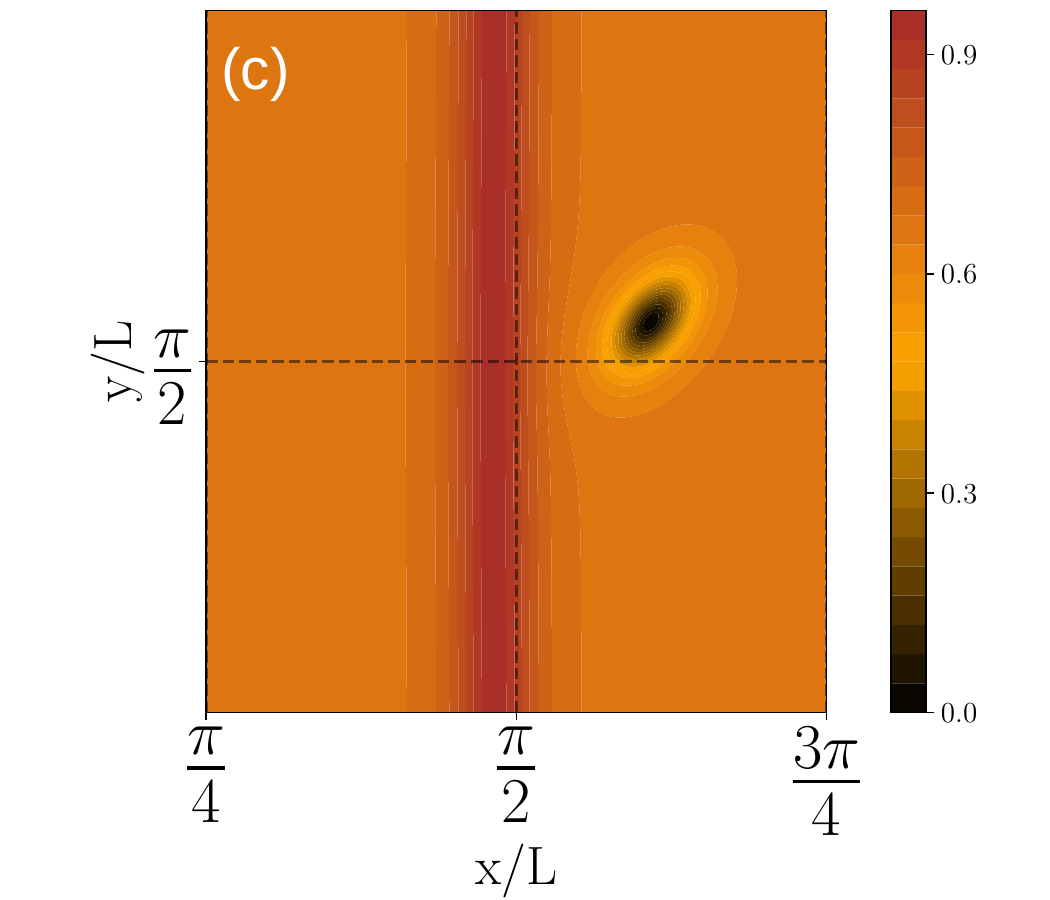}
    }
 \caption{Snapshots of the 
 superconductor order parameter $|\tilde{\psi}|^2$  with $C_4$ symmetry-breaking coupling $\hat{\lambda}_1 = 0.5$ and a repulsive interaction with the wall $\hat{\lambda}_2 = 0.5$. The GL parameter is set as $\kappa = \frac{4}{\sqrt{2}}$.
 The wall width is the same as the superconductor coherence length. As the vortex leaves the wall the trajectory is deviated from the $y = \pi/2$ line, as opposed to the purely biquadratic coupling, where the trajectory evolves along curves with constant $y$.}
  \label{fig:fig4}
  \end{figure*}
\begin{equation}
   \tilde{F}_s \mkern-4mu=\mkern-4mu \int_V \tilde{\alpha}_{GL}|\psi|^2 + \frac{\beta}{2}|\psi|^4 + \frac{\hbar^2}{2m} l_{ij}\mathcal{D}_i\psi  \mathcal{D}_i\psi^* + \frac{(\nabla\times\mathbf{A})^2}{8\pi},
   \nonumber
\end{equation}
where $\tilde{\alpha}_{GL}=\alpha_{GL}+
\lambda_2 \eta^2$ and
\begin{equation}
    l_{ij}=\delta_{ij} + \lambda_1 e_{ij}=
    (1-\lambda_1 \eta)\delta_{ij}+2 \lambda_1 \eta \, n_i n_j .
\end{equation}

After using the expression of $n_i$ with $\alpha=\pi/4$, this becomes
\begin{equation}
    \mathbf{l}=\mathbf{I}_{2x2}+
     \lambda_1 \eta\ \mathbf{\sigma}_1 ,
\end{equation}
with $\mathbf{\sigma}_1$ the symmetric Pauli matrix. The eigenvalues of $\mathbf{l}$ are
\begin{equation}
    l_{\pm}=1 \pm | \lambda_1 \eta| ,
\end{equation}
resulting in elliptical vortices with core size axes
\begin{equation}
    \xi^2_\pm=\frac{\hbar^2 l_\pm}{2m \tilde{\alpha}_{GL}} .
\end{equation}
Defining the eccentricity of an ellipse as $e=(1-a_<^2/a_>^2)^{1/2}$, where $a_>$ ($a_<$) are the larger and smaller axes, we obtain that $e=[2|\lambda_1 \eta|/(1+|\lambda_1 \eta|)]^{1/2}$ (remember that within our approach $|\lambda_1 \eta|<1$, and hence $e<1$).

We next consider the dynamics of an elliptical vortex in the presence of a domain wall. The description in this case is not intuitive as we find the opposite behavior as that observed before for $\lambda_1=0$:
the trajectory of the vortex is not a straight line perpendicular to the wall.
In Fig.~(\ref{fig:fig4}) we present snapshots of the superconductor order parameter for $\lambda_2=0.5$ and $\lambda_1=0.5$, As before, we have chosen for this figure the bare nematic coherence length to be $l_\eta=\xi$. A positive $\hat{\lambda}_2$ causes the vortex to be repelled perpendicular to the wall into a region of increasing nematic order parameter, where the vortex becomes more elliptical as the nematicity increases, as $\hat{\lambda}_1 = 0.5$. This causes the vortex to deviate from its expected perpendicular trajectory.
\begin{figure*}
    \subfloat[\label{fig:fig5a}]{%
      \includegraphics[width=0.32\linewidth]{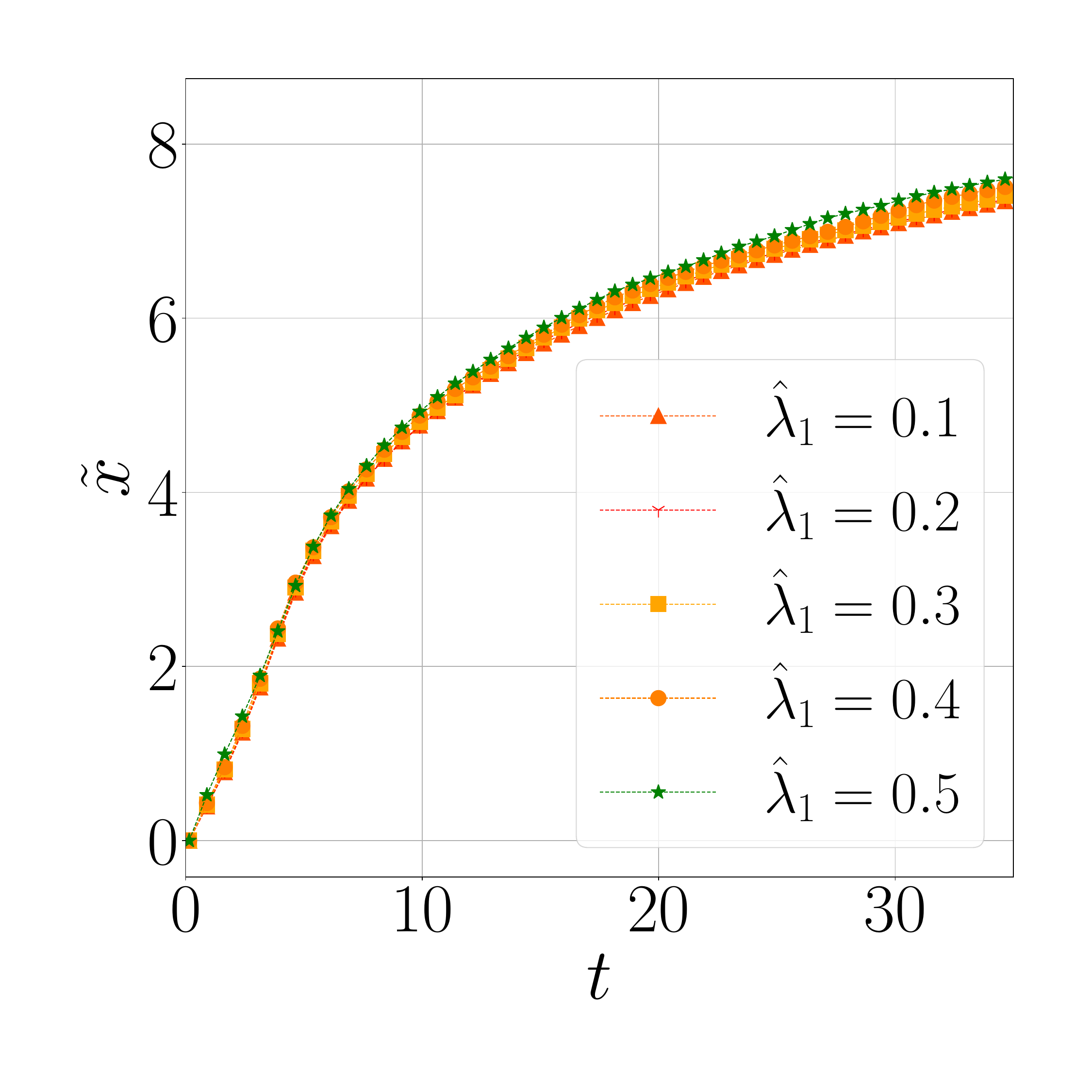}
    }
    \subfloat[\label{fig:fig5b}]{%
      \includegraphics[width=0.32\linewidth]{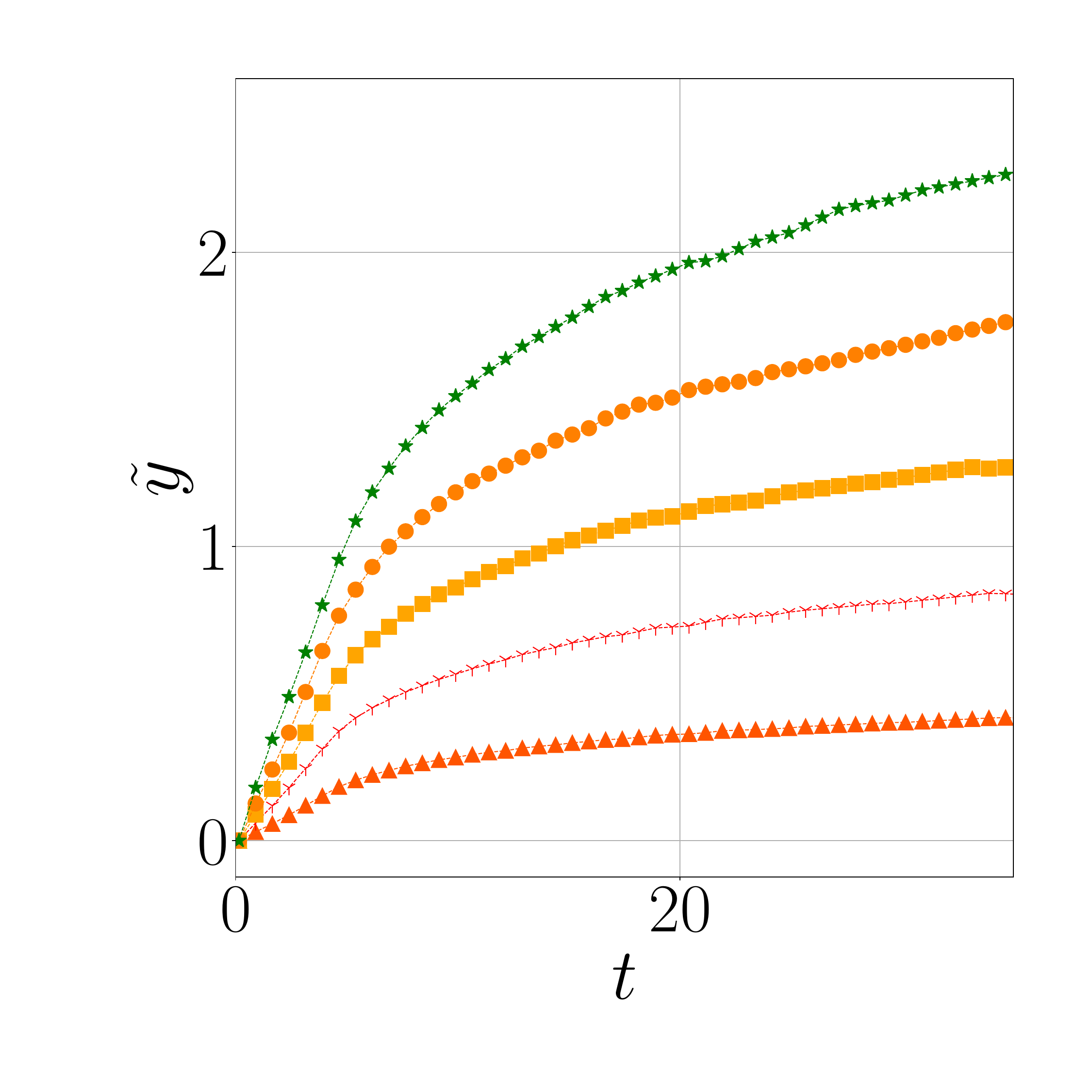}
    }
    \subfloat[\label{fig:fig5c}]{%
      \includegraphics[width=0.32\linewidth]{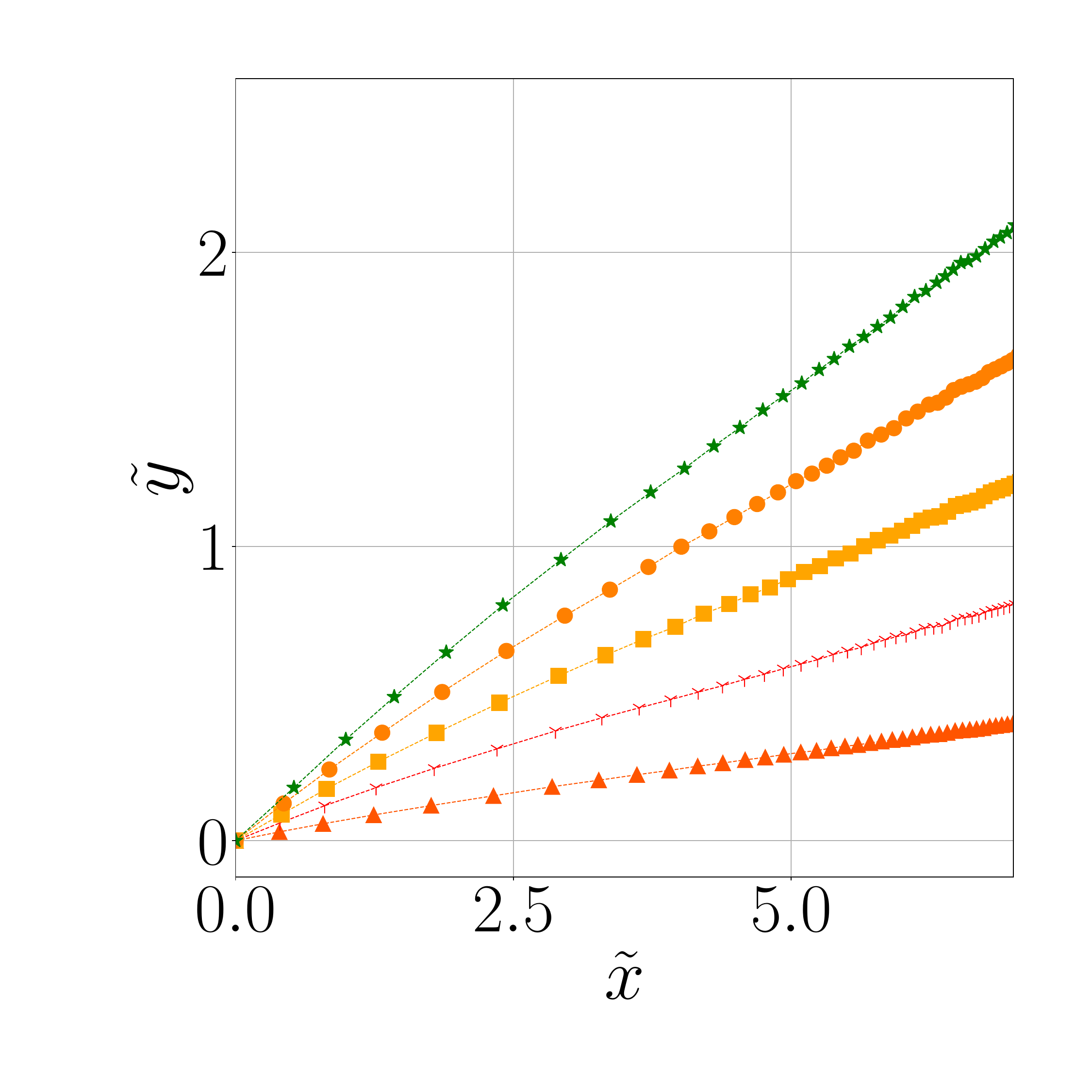}
    }
 \caption{Panels (a) and (b) respectively show $\tilde{x}(t)$ and $\tilde{y}(t)$, measured from the centre of the NDW, as a function of time for different values of $\hat{\lambda}_1$. \textbf{(c)}: Vortex trajectories $\tilde{y}(\tilde{x}(t))$ for several values of $\hat{\lambda}_1$. As the coupling tends to zero, the trajectory becomes perpendicular to the NDW.}
 \label{fig:fig5}
  \end{figure*}

In Fig.~(\ref{fig:fig5}) we show the coordinates $\tilde{x}(t)$, $\tilde{y}(t)$ of the vortex as a function of time, and the trajectories in the plane $\tilde{y}(\tilde{x})$ for fixed $\hat{\lambda}_2$ and varying $\hat{\lambda}_1$.
Contrary to the purely biquadratic case, the vortex trajectory has a component parallel to the wall. While the component perpendicular to the wall is weakly affected by the value of $\hat{\lambda}_1$ (see Fig.~\ref{fig:fig5a}), the component parallel to the wall (see Fig.~\ref{fig:fig5b}) is roughly linear with $\hat{\lambda}_1$.
Plotting the trajectories followed by the vortices (Fig.~\ref{fig:fig5c}) we notice that deviation angles are directly proportional to $\hat{\lambda}_1$ and, as expected, the angle tends to zero when $\lambda_1 \to 0$.

For the attractive case the vortex is pinned to the wall. Due to the different values of the nematic order parameter across the wall, the vortex core displays a peculiar symmetry pattern corresponding to the superposition of two ellipses with axes that are rotated at different sides of the wall, producing a structure with a heart-shaped core. Indeed, as shown  in Fig.~\ref{fig:fig6a}, the vortex core is well described  by the contour lines of the ``heart" function,
\begin{equation}
h(x,y)=(x-x_0)^2+(y-y_0)^2 \pm 2 |\lambda_1 \eta_v||x-x_0| (y-y_0) ,
\end{equation}  
where $(x_0,y_0)=(\pi/2,\pi/2)$ corresponds to the center of the vortex.  This pattern for a vortex core pinned to a wall is indeed very similar to the one observed via STM measurements for FeSe in Ref.~\cite{Watashige_2015}. Note that even though the GL theory considered here is simpler than the one used in Ref.~\cite{Watashige_2015}, the pinned  vortex shape is very similar to that found in theoretical and experimental studies.

\section{Summary and Conclusions}\label{conclusions}
In this paper we addressed vortex-nematic domain wall interactions in a Ginzburg-Landau theory of an $s$-wave superconductor interacting with a real (Ising type) nematic order parameter. The choice of a real order parameter (as opposed to nematic order in liquids described by a tensor order parameter) is inspired by the phenomenology of Fe-based superconductors, where nematicity manifests as the breaking of a $C_4$ symmetry down to a $C_2$ symmetry. Since the nematic order parameter is of Ising-type, it gives rise to domains whose presence is eventually dictated by boundary conditions, thermal history, and additional interactions of the nematic order parameter with other degrees of freedom.

In the simple GL theory we have considered the superconductor order parameter, and the nematic order parameter interaction is introduced via two terms: a biquadratic coupling and a trilinear derivative coupling. The biquadratic term is mainly responsible for the attractive or repulsive character of the interaction, while the trilinear term modifies the in-plane superconductivity, affecting the shape of the vortex and its trajectory (that is not necessarily a straight line perpendicular to the wall in the repulsive case).

In Fe-based superconductors, the nematic domain walls are often linked to twin boundaries separating two different orthorhombic domains. A more complete treatment of the problem would then require the introduction of  elastic terms in the free energy, taking into account the coupling of these elastic terms with the superconductor order parameters and the nematic order parameter. For symmetry reasons, it is expected that twin-boundaries and nematic walls will be strongly coupled. In the simplest scenario it is usually assumed that they are superimposed. Yet their properties could be different. For instance, their width, and as a consequence the extent of its influence on the superconducting properties, could differ. Thus, the final behaviour of the vortex will be the result of these two combined effects, i.e., vortex-nematic domain wall and vortex-twin boundary interaction. 
\begin{figure}
    \subfloat[\label{fig:fig6a}]{%
      \includegraphics[width=0.51\columnwidth]{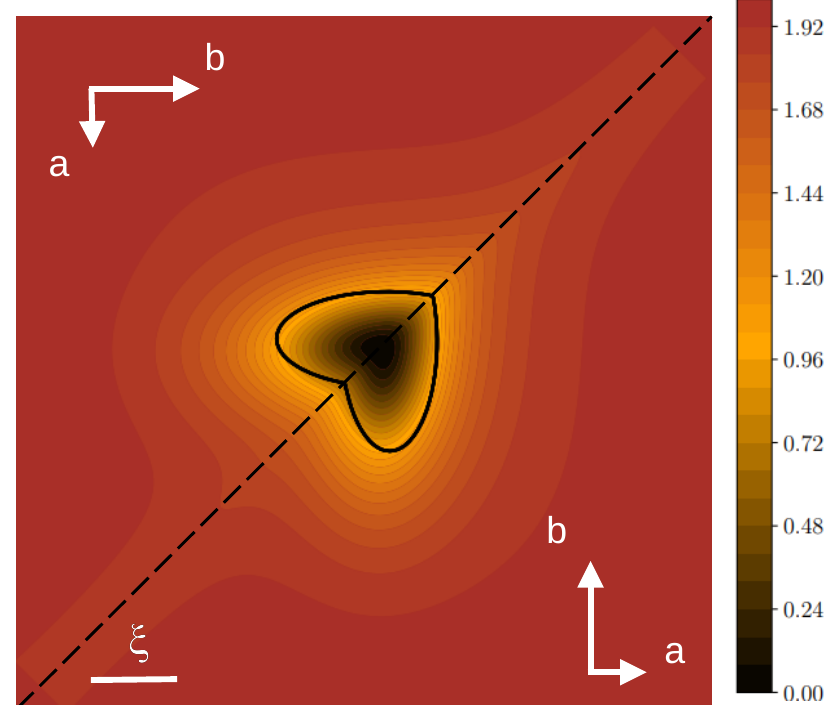}
    }
     \subfloat[\label{fig:fig6b}]{%
      \includegraphics[width=0.44\columnwidth]{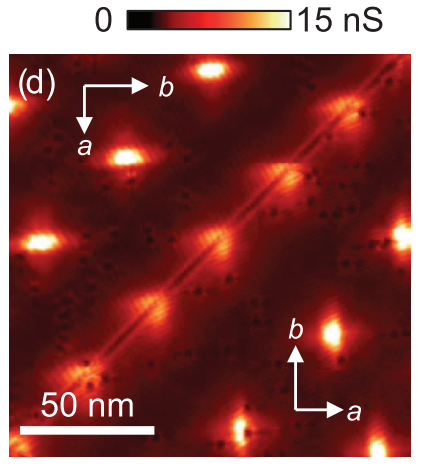}
    }
 \caption{(a) Calculated density of the superconductor order parameter $|\tilde{\psi}|^2$ for a vortex pinned to a nematic domain wall with $\hat{\lambda}_1 = 0.5$ and $\kappa = \frac{4}{\sqrt{2}}$. The nematic domain wall is represented by the black dotted line, and the orthorhombic $a$ and $b$ axes on each domain are shown with white arrows.
 (b) Experimental zero-bias conductance image from T. Watashige \textit{et al.}  \cite{Watashige_2015}, showing vortices pinned to a domain wall at T = 1.5 K and a magnetic field of 1 T applied along the $c$ axis. Crystallographic axes are shown in white arrows. Note that the color code between figures in both panels is inverted: in panel (a) dark corresponds to the vortex core while in (b) the core is represented with white.}        
 \label{fig:fig6}
  \end{figure}

Despite its simplicity, the model captures some relevant features of vortices in Fe based superconductors. As an example, the elongated vortex shape, related in our calculations to the trilinear derivative coupling,  has been observed in FeSe crystalline films (see, e.g, \cite{song}). The attractive vortex-domain wall interaction that results in our model mainly from a negative bi-quadratic coupling term, is consistent with the observed vortex accumulation on FeSe domain boundaries, where degraded superconductivity was found \cite{song,Watashige_2015}. On the other hand, the repulsive character observed in underdoped {\BaFeCoAs},  explained by the observation of enhanced superfluid density on twin boundaries  \cite{Kalisky_2010}, is consistent with a positive bi-quadratic coupling term.
As for the calculated peculiar cordate vortex core shape formed when the vortex is pinned on a domain wall, as shown in Fig.~\ref{fig:fig6a} , it is remarkably similar to the shape observed by the STM measurements shown in Fig.~\ref{fig:fig6b} for cleaved single FeSe crystals \cite{Watashige_2015}. Furthermore, there are some interesting features predicted by the model still to be explored experimentally through direct observation techniques  in Fe-based compounds, as the relationship between the anisotropy of the vortex core, domain wall directions, and vortex trajectories.

Finally, there are several lines of research in which our research can be extended. As mentioned before, the introduction in the formalism of the elastic degrees of freedom is fundamental to understanding the resulting NDM/TB interaction. Finally, the implementation of appropriate boundary conditions to describe vortex lattices, and the way lattices are affected by the presence of domains and the introduction of external currents to study the dynamics of pinning are of interest. We hope to return to these questions in future studies.
 
\section*{Acknowledgements}
RSS, VB, GP and GSL acknowledge support by the University of Buenos Aires, UBACyT 20020170100496BA, Foncyt, PICT Raices -2019-2019-015890, PIP 11220150100653CO and CONICET. PDM acknowledges financial support from UBACYT 20020170100508BA and PICT Grant No. 2018-4298. EF acknowledges support by the US National Science Foundation under the grant DMR-2225920 at the University of Illinois.

\bibliographystyle{unsrt}

\bibliography{refs}

\begin{thebibliography}{10}

\bibitem{kivelson1}
S.~A. Kivelson, E.~Fradkin, and V.~J. Emery.
\newblock Electronic liquid-crystal phases of a doped mott insulator.
\newblock {\em Nature}, 393:550, 1998.

\bibitem{Fradkin_2010}
E.~Fradkin, S.~A. Kivelson, M.~J. Lawler, J.~P. Eisenstein, and A.~P. Mackenzie.
\newblock Nematic fermi fluids in condensed matter physics.
\newblock {\em Annual Review of Condensed Matter Physics}, 1(1):153--178, 2010.

\bibitem{lilly-1999}
M.~P. Lilly, K.~B. Cooper, J.~P. Eisenstein, L.~N. Pfeiffer, and K.~W. West.
\newblock {Evidence for an Anisotropic State of Two-Dimensional Electrons in High Landau Levels}.
\newblock {\em Phys. Rev. Lett.}, 82:394--97, 1999.

\bibitem{kivelson2}
E.~Fradkin and S.~A. Kivelson.
\newblock Liquid-crystal phases of quantum hall systems.
\newblock {\em Phys. Rev. B}, 59:8065, 1999.

\bibitem{borzi-2007}
R.~A. Borzi, S.~A. Grigera, J.~Farrell, R.~S. Perry, S.~J.~S. Lister, S.~L. Lee, D.~A. Tennant, Y.~Maeno, and A.~P. Mackenzie.
\newblock {Formation of a Nematic Fluid at High Fields in {\SROtwo} }.
\newblock {\em Science}, 315:214--217, 2007.

\bibitem{Fradkin_2015}
Eduardo Fradkin, Steven~A. Kivelson, and John~M. Tranquada.
\newblock Colloquium: Theory of intertwined orders in high temperature superconductors.
\newblock {\em Rev. Mod. Phys.}, 87:457--482, 2015.

\bibitem{Fernandes_2022}
R.~F. Fernandes, A.~I. Coldea, H.~Ding, I.~R. Fisher, P.~J. Hirschfeld, and G.~Kotliar.
\newblock Iron pnictides and chalcogenides: a new paradigm for superconductivity.
\newblock {\em Nature}, 601:35, 2022.

\bibitem{Ando_2002}
Yoichi Ando, Kouji Segawa, Seiki Komiya, and A.~N. Lavrov.
\newblock Electrical resistivity anisotropy from self-organized one dimensionality in high-temperature superconductors.
\newblock {\em Phys. Rev. Lett.}, 88:137005, 2002.

\bibitem{Hinkov_2008}
V.~Hinkov, D.~Haug, B.~Fauqué, P.~Bourges, Y.~Sidis, A.~Ivanov, C.~Bernhard, C.~T. Lin, and B.~Keimer.
\newblock Electronic liquid crystal state in the high-temperature superconductor {YBa$_2$Cu$_3$O$_{6.45}$}.
\newblock {\em Science}, 319(5863):597--600, 2008.

\bibitem{Comin_2015}
R.~Comin, R.~Sutarto, E.~H. da~Silva~Neto, L.~Chauviere, R.~Liang, W.~N. Hardy, D.~A. Bonn, F.~He, G.~A. Sawatzky, and A.~Damascelli.
\newblock Broken translational and rotational symmetry via charge stripe order in underdoped \ch{Y Ba_{2} Cu_{3} O_{6+y}}.
\newblock {\em Science}, 347(6228):1335--1339, 2015.

\bibitem{Chuang_2010}
T.-M. Chuang, M.~P. Allan, Jinho Lee, Yang Xie, Ni~Ni, S.~L. Bud’ko, G.~S. Boebinger, P.~C. Canfield, and J.~C. Davis.
\newblock Nematic electronic structure in the "parent" state of the iron-based superconductor \ch{Ca(Fe_{1-x}Co_{x})_{2} As_2}.
\newblock {\em Science}, 327(5962):181--184, 2010.

\bibitem{Prozorov_2009}
R.~Prozorov, M.~A. Tanatar, N.~Ni, A.~Kreyssig, S.~Nandi, S.~L. Bud'ko, A.~I. Goldman, and P.~C. Canfield.
\newblock Intrinsic pinning on structural domains in underdoped single crystals of \ch{Ba(Fe_{1-x} Co_{x})_2 As_{2}}.
\newblock {\em Phys. Rev. B}, 80:174517, 2009.

\bibitem{Kuo_2016}
H.~Kuo, J.~Chu, J.~C. Palmstrom, S.~A. Kivelson, and I.~R. Fisher.
\newblock Ubiquitous signatures of nematic quantum criticality in optimally doped $\ch{Fe}$-based superconductors.
\newblock {\em Science}, 352:958, 2016.

\bibitem{Chu_2010}
Jiun-Haw Chu, James~G. Analytis, Kristiaan~De Greve, Peter~L. McMahon, Zahirul Islam, Yoshihisa Yamamoto, and Ian~R. Fisher.
\newblock In-plane resistivity anisotropy in an underdoped iron arsenide superconductor.
\newblock {\em Science}, 329(5993):824--826, 2010.

\bibitem{Kuo_2012}
H.~H. Kuo, J.~G. Analytis, J.~H. Chu, R.~M. Fernandes, J.~Schmalian, and I.~R. Fisher.
\newblock Magnetoelastically coupled structural, magnetic, and superconducting order parameters in \ch{BaFe_2(As_{1-x} P_x)_2}.
\newblock {\em Phys. Rev. B}, 86:134507, 2012.

\bibitem{Gallais_2013}
Y.~Gallais, R.~M. Fernandes, I.~Paul, L.~Chauvi\`ere, Y.-X. Yang, M.-A. M\'easson, M.~Cazayous, A.~Sacuto, D.~Colson, and A.~Forget.
\newblock Observation of incipient charge nematicity in \ch{Ba(Fe_{1-x}Co_{x})_2 As_{2}}.
\newblock {\em Phys. Rev. Lett.}, 111:267001, 2013.

\bibitem{Tanatar_2016}
M.~A. Tanatar, A.~E. B\"ohmer, E.~I. Timmons, M.~Sch\"utt, G.~Drachuck, V.~Taufour, K.~Kothapalli, A.~Kreyssig, S.~L. Bud'ko, P.~C. Canfield, R.~M. Fernandes, and R.~Prozorov.
\newblock Origin of the resistivity anisotropy in the nematic phase of \ch{FeSe}.
\newblock {\em Phys. Rev. Lett.}, 117:127001, 2016.

\bibitem{Kretzschmar_2016}
F.~Kretzschmar, T.~Böhm, U.~Karahasanovic, B.~Muschler, A.~Baum, D.~Jost, J.´ Schmalian, S.~Caprara, M.~Grilli, C.~Di Castro, J.~G. Analytis, J.-H. Chu, I.~R. Fisher, and R.~Hackl.
\newblock Critical spin fluctuations and the origin of nematic order in \ch{Ba (Fe_{1−x} Co_x)_2 As_{2}}.
\newblock {\em Nature Physics}, 12:560, 2016.

\bibitem{Eckberg-2020}
Chris Eckberg, Daniel~J. Campbell, Tristin metz, John Collini, Halyna Hodovanets, Tyler Drye, Peter Zavalij, Morten~H. Christensen, Rafael~M. Fernandes, Sangjun Lee, Peter Abbamonte, Jeffrey~W. Lynn, and Johnpierre Paglione.
\newblock Sixfold enhancement of superconductivity in a tunable electronic nematic system.
\newblock {\em Nature Physics}, 16:346--350, 2020.

\bibitem{Lee-2021}
Sangjun Lee, John Collini, Stella X.-L. Sun, Matteo Mitrano, Xuefei Guo, Chris Eckberg, Johnpierre Paglione, Eduardo Fradkin, and Peter Abbamonte.
\newblock Multiple charge density waves and superconductivity nucleation at antiphase domain walls in the nematic pnictide {Ba$_{1-x}$Sr$_x$Ni$_2$As$_2$}.
\newblock {\em Phys. Rev. Lett.}, 127:027602, Jul 2021.

\bibitem{Manzeli-2017}
Sajedeh Manzeli, Dmitry Ovchinnikov, Diego Pasquier, Oleg~V. Yazyev, and Andras Kis.
\newblock 2d transition metal dichalcogenides.
\newblock {\em Nature Reviews Materials}, 2:17033, 2017.

\bibitem{Wilson-2023}
Stephen~D. {Wilson} and Brenden~R. {Ortiz}.
\newblock {AV$_3$Sb$_5$ Kagome Superconductors: Progress and Future Directions}.
\newblock {\em arXiv e-prints}, page arXiv:2311.05946, November 2023.

\bibitem{oganesyan-2001}
Vadim Oganesyan, Steven~A. Kivelson, and Eduardo Fradkin.
\newblock Quantum theory of a nematic fermi fluid.
\newblock {\em Phys. Rev. B}, 64:195109, Oct 2001.

\bibitem{halboth-2000}
Christoph~J. Halboth and Walter Metzner.
\newblock $\mathit{d}$-wave superconductivity and pomeranchuk instability in the two-dimensional hubbard model.
\newblock {\em Phys. Rev. Lett.}, 85:5162--5165, Dec 2000.

\bibitem{metzner-2003}
W.~Metzner, D.~Rohe, and S.~Andergassen.
\newblock Soft fermi surfaces and breakdown of fermi-liquid behavior.
\newblock {\em Phys. Rev. Lett.}, 91:066402, Aug 2003.

\bibitem{kivelson-2004}
Steven~A. Kivelson, Eduardo Fradkin, and Theodore~H. Geballe.
\newblock Quasi-one-dimensional dynamics and nematic phases in the two-dimensional emery model.
\newblock {\em Phys. Rev. B}, 69:144505, Apr 2004.

\bibitem{Lv-2009}
Weicheng Lv, Jiansheng Wu, and Philip Phillips.
\newblock Orbital ordering induces structural phase transition and the resistivity anomaly in iron pnictides.
\newblock {\em Phys. Rev. B}, 80:224506, Dec 2009.

\bibitem{Xu-2008}
Cenke Xu, Markus M\"uller, and Subir Sachdev.
\newblock Ising and spin orders in the iron-based superconductors.
\newblock {\em Phys. Rev. B}, 78:020501, Jul 2008.

\bibitem{Fang-2008}
Chen Fang, Hong Yao, Wei-Feng Tsai, JiangPing Hu, and Steven~A. Kivelson.
\newblock Theory of electron nematic order in lafeaso.
\newblock {\em Phys. Rev. B}, 77:224509, Jun 2008.

\bibitem{Nie-2017}
Laimei Nie, Akash~V. Maharaj, Eduardo Fradkin, and Steven~A. Kivelson.
\newblock Vestigial nematicity from spin and/or charge order in the cuprates.
\newblock {\em Phys. Rev. B}, 96:085142, Aug 2017.

\bibitem{Fernandes-2019}
Rafael~M. Fernandes, Peter~P. Orth, and J\"{o}rg Schmalian.
\newblock Intertwined vestigial order in quantum materials: Nematicity and beyond.
\newblock {\em Annual Review of Condensed Matter Physics}, 10(1):133--154, 2019.

\bibitem{Chu2012}
Jiun-Haw Chu, Hsueh-Hui Kuo, James~G. Analytis, and Ian~R. Fisher.
\newblock Divergent nematic susceptibility in an iron arsenide superconductor.
\newblock {\em Science}, 337(6095):710--712, 2012.

\bibitem{Kalisky_2010}
B.~Kalisky, J.~R. Kirtley, J.~G. Analytis, Jiun-Haw Chu, A.~Vailionis, I.~R. Fisher, and K.~A. Moler.
\newblock Stripes of increased diamagnetic susceptibility in underdoped superconducting $\ch{Ba(Fe_{1-x} Co_{x})_{2} As_{2}}$ single crystals: Evidence for an enhanced superfluid density at twin boundaries.
\newblock {\em Phys. Rev. B}, 81:184513, 2010.

\bibitem{sanches_nature}
J.~J. Sanchez, P.~Malinowski, J.~Mutch, J.~Liu, J.~W. Kim, P.~J. Ryan, and J.~Chu.
\newblock The transport–structural correspondence across the nematic phase transition probed by elasto x-ray diffraction.
\newblock {\em Nat. Mater.}, 20:1519, 2021.

\bibitem{Bartlett_2021}
J.~M. Bartlett, A.~Steppke, S.~Hosoi, H.~Noad, J.~Park, C.~Timm, T.~Shibauchi, A.~P. Mackenzie, and C.~W. Hicks.
\newblock Relationship between transport anisotropy and nematicity in \ch{FeSe}.
\newblock {\em Phys. Rev. X}, 11:021038, 2021.

\bibitem{Blatter1994}
G.~Blatter, M.~V. Feigel'man, V.~B. Geshkenbein, A.~I Larkin, and V.~M. Vinokur.
\newblock Vortices in high-temperature superconductors.
\newblock {\em Rev. Mod. Phys.}, 66:1125, 1994.

\bibitem{Kwok_1996}
W.~K. Kwok, J.~A. Fendrich, V.~M. Vinokur, A.~E. Koshelev, and G.~W. Crabtree.
\newblock Vortex shear modulus and lattice melting in twin boundary channels of \ch{Y Ba_{2} Cu_{3} O_{7-$\delta$}}.
\newblock {\em Phys. Rev. Lett.}, 76:4596--4599, 1996.

\bibitem{Crabtree_1996}
G.W. Crabtree, G.K. Leaf, H.G. Kaper, V.M. Vinokur, A.E. Koshelev, D.W. Braun, D.M. Levine, W.K. Kwok, and J.A. Fendrich.
\newblock Time-dependent ginzburg-landau simulations of vortex guidance by twin boundaries.
\newblock {\em Physica C: Superconductivity}, 263(1):401--408, 1996.
\newblock Proceedings of the International Symposium on Frontiers of High - Tc Superconductivity.

\bibitem{Herbsommer_2000}
J.~A. Herbsommer, G.~Nieva, and J.~Luzuriaga.
\newblock Interplay between pinning energy and vortex interaction in $\ch{Y Ba_2 Cu_{3} 0_{7-$\delta$}}$ with oriented twin boundaries in tilted magnetic fields: Bitter decoration and tilt-modulus measurements.
\newblock {\em Phys. Rev. B}, 62:3534--3541, 2000.

\bibitem{Marziali_2013}
M.~Marziali~Berm\'udez, G.~Pasquini, S.~L. Bud'ko, and P.~C. Canfield.
\newblock Correlated vortex pinning in slightly orthorhombic twinned $\ch{Ba (Fe_{1-x} Co_{x})_2 As_{2}}$ single crystals: Possible shift of the vortex-glass/liquid transition.
\newblock {\em Phys. Rev. B}, 87:054515, 2013.

\bibitem{schmidt}
J.~Schmidt, V.~Bekeris, G.~S. Lozano, M.~V. Bortule, M.~Marziali Bermudez, C.~W. Hicks, P.~C. Canfield, E~Fradkin, and G.~Pasquini.
\newblock Nematicity in the superconducting mixed state of strain detwinned underdoped $\ch{Ba (Fe_{1-x} Co_{x})_2 As_{2}}$.
\newblock {\em Phys. Rev.B}, 99:064515, 2019.

\bibitem{song}
C.~L. Song, Y.~L. Wang, P.~Cheng, Y.~P. Jiang, W.~Li, T.~Zhang, Z.~Li, K.~He, L.~Wang, and J.~F.~Jia et~al.
\newblock Direct observation of nodes and twofold symmetry in fese superconductor.
\newblock {\em Science}, 332:1410, 2011.

\bibitem{songprl}
Can-Li~Song et~al.
\newblock Suppression of superconductivity by twin boundaries in \ch{FeSe}.
\newblock {\em Phys.Rev.Lett.}, 109:137004, 2012.

\bibitem{Watashige_2015}
T.~Watashige, Y.~Tsutsumi, T.~Hanaguri, Y.~Kohsaka, S.~Kasahara, A.~Furusaki, M.~Sigrist, C.~Meingast, T.~Wolf, H.~v. L\"ohneysen, T.~Shibauchi, and Y.~Matsuda.
\newblock Evidence for time-reversal symmetry breaking of the superconducting state near twin-boundary interfaces in $\ch{FeSe}$ revealed by scanning tunneling spectroscopy.
\newblock {\em Phys. Rev. X}, 5:031022, 2015.

\bibitem{zhang}
Irene P.~Zhang et~al.
\newblock Imaging anisotropic vortex dynamics in \ch{FeSe}.
\newblock {\em Phys.Rev.}, 100:024514, 2019.

\bibitem{yin}
Yi~Yin et~al.
\newblock Scanning tunneling spectroscopy and vortex imaging in the iron pnictide superconductor \ch{Ba Fe_{1.8} Co_{0.2} As_{2}}.
\newblock {\em Phys.Rev.Lett.}, 102:097002, 2009.

\bibitem{Kalisky_2011}
B.~Kalisky, J.~R. Kirtley, J.~G. Analytis, J.-H. Chu, I.~R. Fisher, and K.~A. Moler.
\newblock Behavior of vortices near twin boundaries in underdoped $\ch{Ba (Fe_{1-x} Co_{x})_2 As_{2}}$.
\newblock {\em Phys. Rev. B}, 83:064511, 2011.

\bibitem{Yagil_2016}
A.~Yagil, Y.~Lamhot, A.~Almoalem, S.~Kasahara, T.~Watashige, T.~Shibauchi, Y.~Matsuda, and O.~M. Auslaender.
\newblock Diamagnetic vortex barrier stripes in underdoped $\ch{Ba Fe_2 (As_{1-x} P_x)_2}$.
\newblock {\em Phys. Rev. B}, 94:064510, 2016.

\bibitem{Hecher_2018}
J.~Hecher, S.~Ishida, D.~Song, H.~Ogino, A.~Iyo, H.~Eisaki, M.~Nakajima, D.~Kagerbauer, and M.~Eisterer.
\newblock Direct observation of in-plane anisotropy of the superconducting critical current density in $\ch{Ba (Fe_{1-x} Co_{x})_2 As_{2}}$ crystals.
\newblock {\em Phys. Rev. B}, 97:014511, 2018.

\bibitem{hashimoto}
H.~Q. Yamamotom et~al. T.~Hashimoto, Y.~Ota.
\newblock Superconducting gap anisotropy sensitive to nematic domains in fese.
\newblock {\em Nature Communications}, 9:282, 2018.

\bibitem{ren}
Zhen~Reng et~al.
\newblock Nanoscale decoupling of electronic nematicity and structural anisotropy in fese thin films.
\newblock {\em Nature Communications}, 12:10, 2021.

\bibitem{chowdhury}
D.~Chowdhury, E.~Berg, and S.~Sachdev.
\newblock Nematic order in the vicinity of a vortex in superconducting \ch{FeSe}.
\newblock {\em Phys. Rev.B}, 84:205113, 2011.

\bibitem{us}
R~Severino et~al.
\newblock Vortices in a ginzburg-landau theory of superconductors with nematic order.
\newblock {\em Phys. Rev. B}, 106:094512, 2022.

\bibitem{schmid}
A.~Schmid.
\newblock A time dependent ginzburg-landau equation and its application to the problem of resistivity in the mixed state.
\newblock {\em Phys. Kondens Materie}, 5:302, 1966.

\bibitem{ghosta}
P.~D. Mininni, D.~Rosenberg, R.~Reddy, and A.~Pouquet.
\newblock A hybrid mpi–openmp scheme for scalable parallel pseudospectral computations for fluid turbulence.
\newblock {\em Parallel Computing}, 37:316, 2011.

\bibitem{ghostb}
D.~Rosenberg, P.~D. Mininni, R.~Reddy, and A.~Pouquet.
\newblock Gpu parallelization of a hybrid pseudospectral geophysical turbulence framework using cuda.
\newblock {\em Atmosphere}, 11:178, 2020.

\bibitem{nore}
C.~Nore, M.~Abid, and M.~E. Brachet.
\newblock Decaying kolmogorov turbulence in a model of superflow.
\newblock {\em Physics of Fluids}, 9:2644, 1997.

\end{thebibliography}

\end{document}